\documentclass[preprint*]{JHEP3} 

\usepackage{epsfig,multicol,bbm}
\usepackage{amsmath}
\usepackage{amsfonts}
\usepackage{amssymb}
\usepackage{graphicx}
\input amssym.def
\input amssym

\newcommand\fverb{\setbox\fverbbox=\hbox\bgroup\verb}
\newcommand\fverbdo{\egroup\medskip\noindent%
			\fbox{\unhbox\fverbbox}\ }
\newcommand\fverbit{\egroup\item[\fbox{\unhbox\fverbbox}]}
\newbox\fverbbox

\newcommand{\be}{\begin{equation}}
\newcommand{\ee}{\end{equation}}

\title{The Universe as a Set of Topological Fluids with Hierarchy and Fine Tuning of Coupling Constants in Terms of Graph Manifolds}

\author{Vladimir N. Efremov\\ Mathematics Department, CUCEI, University of Guadalajara,\\ Av. Revoluci\'on, 1500, C.P. 44430, Guadalajara, Mexico\\ E-mail: \email{efremov@cencar.udg.mx}}
\author{Alfonso M. Hern\'andez Magdaleno\\ Mathematics Department, CUCEI, University of Guadalajara,\\ Av. Revoluci\'on, 1500, C.P. 44430, Guadalajara, Mexico\\ E-mail: \email{137mag@gmail.com}}
\author{Fernando I. Becerra Lopez\\ Mathematics Department, CUCEI, University of Guadalajara,\\ Av. Revoluci\'on, 1500, C.P. 44430, Guadalajara, Mexico\\ E-mail: \email{mail@ferdx.com}}

\received{~} 		
\accepted{~}		

\abstract{The hierarchy and fine tuning of the gauge coupling constants are described 
on the base of topological invariants (Chern classes interpreted as filling factors) characterizing a collection of fractional topological fluids emerging from three dimensional graph manifolds, which play the role of internal spaces in the Kaluza-Klein approach to the topological BF theory. The hierarchy of BF gauge coupling constants is simulated by diagonal elements and eigenvalues of rational linking matrices of tree graph manifolds pasted together from Brieskorn (Seifert fibered) homology spheres. Specific examples of graph manifolds are presented which contain in their linking matrices the hierarchy of coupling constants distinctive for the dimensionless coupling constants in our Universe. The fine tuning effect is simulated owing to the special numerical properties of diagonal elements of the linking matrices. We pay a particular attention to fine tuning problem for the cosmological constant.}

\keywords{Field Theories in Higher Dimensions; Topological Field Theories}

\begin{document}

\section{INTRODUCTION} \label{s1} 
\setcounter{equation}{0}
The different aspects of gauge hierarchy and fine tuning in particle physics and cosmology are important conceptual problems of modern physics. We try to solve them on the base of exclusive topological properties of graph manifolds built out of Brieskorn (Seifert fibered) homology spheres by means of a topology operation, known as plumbing. Two rather new empirical dates sharpen the interest and inspirit the reconsideration of hierarchy and fine tuning problems, as well as their interconnections. The first one is the discovery in 1998 (by measuring the apparent luminosity of distant supernovae \cite{{Riess},{Perlmutter}}) of the accelerated expansion of the Universe, consistent with a positive cosmological constant \cite{{TegAgReWil},{Bous}}
\be\label{1.1}\rho_{\Lambda} = (1.35 \pm 0.15)\times 10^{-123},\ee
and inconsistent with $\rho_{\Lambda}=0$. More recent cross-checks corroborated this 
conclusion. The second experiment advance, consisting in the observation of Higgs boson like particle with the mass of 126 GeV \cite{{Aad},{Chatrchyan}}, suggests (together with non-discovery of SUSY particles at the LHC) that the SUSY breaking scale is considerably higher than the electroweak scale. This already raises doubt of the low energy SUSY as a solution to the hierarchy problem (see, for example, \cite{1301.1137}).

We consider the gauge hierarchy problem in the aspect of hierarchy of dimensionless coupling  constants including the cosmological constant (see the first column and Notes in table 2 for the list of dimensionless low-energy coupling constants based on \cite{TegAgReWil}). The problem of cosmological constant (vacuum energy density), as a rule, stands apart and it is regarded as a fine tuning problem since quantum field theory predicts $|\rho_{\Lambda}|$ to be some 60 to 120 orders of magnitude larger than the experimental bound (\ref{1.1}) (see, for example, \cite{{Rubakov}, {Bous}} and section 5 of this paper for review of several known contributions to $\rho_{\Lambda}$). There are different approaches to the cosmological constant problem such as the model of ``eternal inflation'' \cite{Linde}, the concept of ``string landscape'' \cite{BousPolchin}, the ``baby universes model'' \cite{{LavRubTin}, {GiddStrom}, {Colem}}, {\it et cetera}. But by these methods one cannot predict and calculate the real value of the cosmological constant, as well as cannot explain (outside of the anthropic principle) the existence of such a small vacuum energy scale as (\ref{1.1}) dictates. This aspect of the problem is in some sense analogous to the hierarchy problem.

At the same time there has been developed an excellent theoretical description of the hierarchical systems such as fractional quantum Hall states \cite{{WenObzor},{WenZee}} and fractional topological insulators \cite{{ChoMoore},{DiamantII},{MacQiKarZang}}, which are based on the existence of a new state of matter characterized by a new type of order: topological order. These methods are applied not only to the description of three dimensional phenomena such as quantum Hall effect, but also for explanation of four dimensional systems connected with a topological superconductivity \cite{Diamant05} or a topological confinement \cite{DiamantI}, without any spontaneous symmetry breaking pattern. One of the basic characteristics of topological order is that the long-distance effective field theory for such states involves a topological field theory \cite{BirmBlauRak}. In (2+1)-dimensional case the dominant part of gauge action for low-energy topological quantum Hall fluids is Chern-Simons term. The effective description of hierarchical fractional quantum Hall system with multiple $U(1)$ gauge fields $A^I$ (I=1,\dots R) is realized by the following action (see \cite{{WenKniga}, {FujitaLi}} for details):
\be\label{1.2} S=\frac{1}{4\pi} \int_{M^3}\left( \sum_{I,J=1}^R K_{IJ} A^I\wedge dA^{J} + 2\sum_{I=1}^R t_I A_{\rm ext}\wedge d A^{I}\right), \ee
where $K_{IJ}$ is an integer tridiagonal matrix ($K$-matrix), $t_I$  is called the charge vector and $A_{\rm ext}$ is a external or background potential. The rational filling factor which determines the Hall conductivity is defined as $\nu:=t_I K^{IJ}t_J$, where $K^{IJ}$ is the matrix inverse to $K_{IJ}$. It is possible to interpret the rational matrix $K^{IJ}$ as coupling constant matrix for the collection of Abelian fields $\{ A^I\}$, which describe the set of quasi-particles corresponding to the different condensates (condensate over another condensate, see \cite{WenObzor} for details).

Recently, a new topological fluids in three spatial dimensions have attracted intense attention. We already have mentioned about the so-called topological insulators \cite{{ChoMoore}, {HasanMoore}} as well as the dynamical systems describing topological superconductivity \cite{Diamant05} and topological confinement \cite{DiamantI}.
All these topological fluids are characterized by low-energy {\it effective} actions containing the topological BF terms \cite{BirmBlauRak}, which are the higher-dimensional generalization of Chern-Simons term. It is possible to introduce 
the hierarchical structure in the BF-model by means of Kaluza-Klein procedure that induces a set of four-dimensional Abelian gauge potentials $A^I$ with coupling constants matrix analo\-gous to $K$-matrix characterizing the hierarchical fractional quantum Hall effect. We have realized a BF-model of this type in section 3 of this paper. To obtain a specific hierarchy of coupling constants that exists in our universe we have used special type of internal spaces in Kaluze-Klein approach, namely graph manifolds (see  reviews in \cite{{Neum77}, {NeumLect}}), which consist of the finite number of Seifert fibered pieces, so-called JSJ-pieces, corresponding to the decomposition of three-dimensional manifolds investigated by Jaco, Shalen \cite{JaSha} and Johannson \cite{Joh}, known as JSJ-decomposition.
The BF theory may be also obtained as a result of the Julia-Toulouse approach  \cite{JulToul} which determines the effective model describing the regime of Abelian gauge theory with condensed topological defects \cite{Braz1202.3798}.  
This mechanism may lead to a more physical interpretation of our rather formal results.
Our hope is based on the fact that
the dynamical equations which we obtain from a formal Kaluza-Klein approach, are of the same type as the classical equations obtained in \cite{BrazJHEP} from the Julia-Toulouse approach in the case when generalized
``electric'' defects condensate.

Note that the interest to the Seifert fibered manifolds appears in the last decade and is connected with the reconsideration of Chern-Simons gauge theory on the Seifert manifolds possessing  $U(1)$-invariant contact structures (in general a $U(1)$-action is pseudofree \cite{FintStPsOr}), which permit to apply the non-abelian localization \cite{BeasleyWitt}.  Moreover, the Seifert fibered manifolds, in particular lens spaces, are very useful to investigate the contribution of fractional instantons to the black hole partition functions \cite{GriTan}.
We use the calculation methods developed in the last two cited articles to calculate the rational linking matrices of the graph manifolds in the section 2.2. These linking matrices play the role of $K$-matrices and contain the information, not only about hierarchy of coupling constants, but also about fine tuning phenomenon which is universal for all gauge couplings involved in our model.

The paper is organized as follows. In subsection 2.1 we review the JSJ-decomposition of plumbing graph manifolds and define a new concept of JSJ-covering. Moreover we calculate the linking (intersection) numbers for different Seifert fibered structures defined on lens spaces. In subsection 2.2 the rational linking matrices of plumbing  graph manifolds are expressed via the sum of continued fractions which are interpreted as filling factors corresponding to the collection of hierarchical topological fluids, putting together according to the structure of tree plumbing graph. Further we show that these rational linking matrices coincide with the reduced plumbing matrices, defined by W. Neumann in \cite{Neum77}, and obtained by partial (Gauss) diagonalization of integer linking matrix of graph manifold.
An interpretation in terms of ``baby universes'' is given for the result of this partial diagonalization.

In the section 3 the seven-dimensional BF-model is constructed, which contains the hierarchical structure typical for fractional quantum Hall fluids and topological mass generation. In this model it is possible to interpret the linking matrices of graph manifolds as BF coupling constants matrices. Also we try to read our results in terms of Julia-Toulouse approach.

In subsection 4.1 the principal ensemble of Brieskorn homology spheres (Bh-spheres) with the simplest Seifert structures is defined. These Bh-spheres play the role of building blocks for the graph manifolds forming internal spaces in Kaluza-Klein approach described in the section 3. In this subsection we also give the simplest example of modeling a fine tuning effect by means of Bh-spheres. In subsection 4.2 we complete the construction of the set of plumbing graph manifolds, which contain in their linking matrices the hierarchy of dimensionless coupling constants of gauge interactions ``switched on'' in our universe.

In section 5 we give the results of numerical calculations demonstrating that the defined set of linking matrices formally simulates the fine tuning effect and hierarchy of (at least) low energy empirical coupling constants of the fundamental interactions including the cosmological one. 

In section 6 we discuss the main results of our paper.    

The standard notations ${\mathbb Z}$ and ${\mathbb Q}$  are used for the sets of integer and rational numbers, respectively. The symbol ${\mathbb N}$ (${\mathbb Z}^+$) denotes the set of positive (nonnegative) integers and ${\mathbb Z}_p={\mathbb Z}/p{\mathbb Z}$ is a cyclic group of order $p\in {\mathbb N}$, $p>1$.

\section{Graph Manifolds and Rational Linking Matrices} \label{s2} \setcounter{equation}{0}

Topologically non-trivial manifolds, which we shall use in this paper as the internal spaces in Kaluza-Klein approach, pertain to the famous class of three-dimensional graph manifolds \cite{{Neum77}, {SavB2}}. From the point of view of Jaco-Shalen-Johannson  classification \cite{{JaSha}, {Joh}}, a graph manifold is a three-dimensional space which has only Seifert fibered pieces $M^I$, $I=1,\dots, R$  in its decomposition along a set of incompressible tori (JSJ-decomposition). 
Let each piece $M^I$ be a Seifert fibered three-manifold  (with boundary) characterized by the unnormalized Seifert invariants $\{ (a_1^I, b_1^I),\dots,(a_{k_I}^I, b_{k_I}^I)\}$ (see Appendix). Recall \cite{{FurutaSteer}, {BeasleyWitt}} that $M^I$ may be represented as a U(1)-bundle associated with a line $V$-bundle over orbifold $\hat{\Sigma^I}$ with conical singularities at the marked points with cone angles $2\pi/a_i^I$, $i=1,\dots,k_I$. We consider only tree type graph manifolds which are the result of Seifert homology spheres plumbing, so the genus of the orbifold is always zero ($g^I=0$). Each Seifert fibered piece $M^I$ is characterized by the rational Euler number 
\be\label{2.1}  e^I=-\sum_{i=1}^{k_I} b_i^I/a_i^I. \ee
W. Neumann in  \cite{Neum77} defined the reduced plumbing matrix of the toral JSJ-decomposition as follows: $K^{II}_{\rm reduced}= e^I$ and $K^{IJ}_{\rm reduced}= 1/p^{IJ}$, where $p^{IJ}$ is the fiber intersection number over the torus that separates $M^I$ from $M^J$. In the same paper it was shown that this matrix appears as a result of partial diagonalization of the integer linking matrix of the graph manifold. In this section we shall demonstrate that for a tree graph manifold the reduced plumbing matrix coincides with the rational linking matrix $K^{IJ}$ (defined in Subsection \ref{s2.2}), 
which we shall interpreted as the gauge coupling constants matrix in the topological BF-model constructed in the Section 3. Also we shall observe that each diagonal element $K^{II}$ can be represented as a sum of filling factors of topological fluids, associated with the node $N^I$ of the graph $\Gamma_p$.

\subsection{JSJ-coverings of plumbing graph manifolds} \label {s2.1}

We begin from the definition of plumbing graph $\Gamma_p$ as a finite one-dimensional simplicial complex, which does not contain multiple edges and loops, that is several edges from one vertex to another or edges connecting a vertex to itself, {\it i.e.} we consider only the graph of tree type. An integer weight $e_i$ is assigned to each vertex of $\Gamma_p$, this weight is known as the Euler number of the principal $S^1$-($U(1)$-)bundle, corresponding to $i$-th vertex $v_i$. To understand plumbing in more detail, we need a precise description of the "simplest" bundles involved. If $A=\left(
\begin{array}{ll}
a & b\\
c & d
\end{array}
\right)\in SL(2,\mathbb{Z})$, we denote also by $A$ the diffeomorphism of the torus $T^2=S^1\times S^1$ given by $A:S^1\times S^1\rightarrow S^1\times S^1$, $A(t_1,t_2)=(t_1^a t_2^b,t_1^c t_2^d)$ in the multiplicative form. Writing $t_1=e^{i2\pi x}$, $t_2=e^{i2\pi y}$ we can represent this action in additive form: \be \label{2.5} \left(
\begin{array}{ll}
a & b\\
c & d
\end{array}
\right)
\left(
\begin{array}{l}
x\\
y
\end{array}
\right)=
\left(
\begin{array}{l}
ax+by\\
cx+dy
\end{array}
\right), \ee 
where effectively $x,y\in \mathbb{R}/\mathbb{Z}$ \cite{Hirz}.
We define the bundle $M(e_i)$ associated to each vertex $v_i$ as $S^1$-bundle over $S^2$ with the Euler number $e_i$, which can be pasted together from two trivial bundles over $D^2$ as follows \cite{{Hirz}, {NeumLect}} \be \label{2.6}
D^2\times S^1 \cup_{H_i} D^2\times S^1, H_i: \partial(D^2\times S^1)=S^1\times S^1 \rightarrow S^1\times S^1=\partial(D^2\times S^1) \ee where \be \label{2.7}
H_i
\left(
\begin{array}{l}
x\\
y
\end{array}
\right)=
\left(
\begin{array}{ll}
-1 & 0\\
-e_i & 1
\end{array}
\right)
\left(
\begin{array}{l}
x\\
y
\end{array}
\right)=
\left(
\begin{array}{l}
-x\\
y-e_ix
\end{array}
\right). \ee
Note that the above is a well known description of the lens space $L(e_i,1)$, so the total space of the bundle is $M(e_i)=L(e_i,1)$.\\

\noindent{\bf Remark 2.1.}
The total space $M(e_i)$ of the principal $S^1$-bundle (which we denote in the same way) can be identified as the boundary of the associated $2$-disc (line) bundle over $S^2$, which is the bundle we need for plumbing of 4-dimensional graph manifolds. However, in many cases it is more natural use directly the graph manifold $M(\Gamma_p)=\partial P(\Gamma_p)$ rather than $P(\Gamma_p)$.\\

To perform pasting operation, which is known as {\it plumbing} between the $S^1$-bundles, we must use the trivial bundles over annuli 
\be \label{2.8}M_A(e_i)=A\times S^1 \cup_{H_i} A\times S^1 \ee
where A is an annulus or twice punctured sphere $S^2$.
The manifold $M(\Gamma_p)$ is pasted together from the manifolds $M_A(e_i)$ as follows \cite {NeumLect} : whenever vertices $v_i$ and $v_j$ are connected by an edge $\sigma_{ij}$ in $\Gamma_p$ we paste a boundary component $S^1\times S^1$ of $M_A(e_i)$ to a boundary component $S^1\times S^1$ of $M_A(e_j)$ by the map $J=\left(
\begin{array}{ll}
0 & 1\\
1 & 0
\end{array}
\right)$: \be \label{2.9}
A\times S^1 \cup_{H_i} A\times S^1 \cup_J A\times S^1 \cup_{H_j} A\times S^1, \ee
so the base and fiber coordinates are exchanged under the plumbing operation.
For three-dimensional manifolds this operation is known also as a splicing \cite{EisNeum}.
Thus the edge $\sigma_{ij}$ corresponds to the torus $T^2_{ij}=S^1\times S^1$ along which the pieces $M_A(e_i)$ and $M_A(e_j)$ pasted together.

For example, suppose we have a maximal \footnote{The chain is maximal if it cannot be included in some larger chain} internal chain  of length $k$, $\{v_1,...,v_k\}$ with Euler numbers $\{e_1,...,e_k\}$ embedded in a tree graph $\Gamma_p$ as in Fig. \ref{f1}:

\begin{figure}[h]
\begin{center}
\setlength{\unitlength}{1pt}
\begin{picture}(250,70)
\put(0,35){\line(1,-1){30}}
\multiput(0,-10)(0,10){3}{.}
\put(0,-35){\line(1,1){30}}
\put(32,0){\circle{10}}
\put(38,0){\line(1,0){20}}
\put(58,-0.5){\makebox(.5,.5){$\bullet$}}
\put(58,0){\line(1,0){20}}
\put(78,-0.5){\makebox(.5,.5){$\bullet$}}
\put(78,0){\line(1,0){20}}
\multiput(108,-0.5)(10,0){3}{.}
\put(138,0){\line(1,0){20}}
\put(158,-0.5){\makebox(.5,.5){$\bullet$}}
\put(158,0){\line(1,0){20}}
\put(183,0){\circle{10}}
\put(185,5){\line(1,1){30}}
\multiput(215,-10)(0,10){3}{.}
\put(185,-5){\line(1,-1){30}}

\put(32,15){\makebox(.5,.5){$e^I$}}
\put(32,-15){\makebox(.5,.5){$N^I$}}
\put(58,10){\makebox(.5,.5){$e_1$}}
\put(58,-10){\makebox(.5,.5){$v_1$}}
\put(78,10){\makebox(.5,.5){$e_2$}}
\put(78,-10){\makebox(.5,.5){$v_2$}}
\put(158,10){\makebox(.5,.5){$e_k$}}
\put(158,-10){\makebox(.5,.5){$v_k$}}
\put(185,15){\makebox(.5,.5){$e^J$}}
\put(182,-15){\makebox(.5,.5){$N^J$}}

\end{picture}\\
\vspace*{1.cm}
\caption{A maximal internal chain of length k embedded in a tree graph $\Gamma_p$. \label{f1}}
\end{center}
\end{figure}
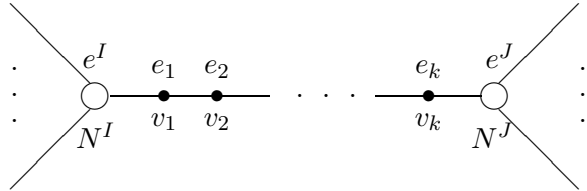
The plumbing according to this portion of $\Gamma_p$ thus gives the pasting \be \label{2.10}
M(N^I) \cup_J (A\times S^1 \cup_{H_i} A\times S^1) \cup_J \cdots \cup_J (A\times S^1 \cup_{H_k} A\times S^1) \cup_J M(N^J), \ee
where $M(N^I)$ is a $S^1$-bundle corresponding to the node $N^I$.

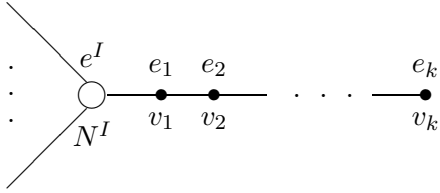
\begin{figure}[h]
\begin{center}

\setlength{\unitlength}{1pt}
\begin{picture}(160,70)
\put(0,35){\line(1,-1){30}}
\multiput(0,-10)(0,10){3}{.}
\put(0,-35){\line(1,1){30}}
\put(32,0){\circle{10}}
\put(38,0){\line(1,0){20}}
\put(58,-0.5){\makebox(.5,.5){$\bullet$}}
\put(58,0){\line(1,0){20}}
\put(78,-0.5){\makebox(.5,.5){$\bullet$}}
\put(78,0){\line(1,0){20}}
\multiput(108,-0.5)(10,0){3}{.}
\put(138,0){\line(1,0){20}}
\put(158,-0.5){\makebox(.5,.5){$\bullet$}}

\put(32,15){\makebox(.5,.5){$e^I$}}
\put(32,-15){\makebox(.5,.5){$N^I$}}
\put(58,10){\makebox(.5,.5){$e_1$}}
\put(58,-10){\makebox(.5,.5){$v_1$}}
\put(78,10){\makebox(.5,.5){$e_2$}}
\put(78,-10){\makebox(.5,.5){$v_2$}}
\put(158,10){\makebox(.5,.5){$e_k$}}
\put(158,-10){\makebox(.5,.5){$v_k$}}

\end{picture}\\
\vspace*{1.cm}
\caption{A maximal terminal chain of length k embedded in a tree graph $\Gamma_p$. \label{f2}}
\end{center}
\end{figure}

It is appropriate here to define more exactly certain notions \cite{PaPoPa}.
Vertices with at least three edges are called nodes. We shall use plumbing graphs with nodes of minimal valence (n=3) only. This type of graph corresponds to plumbing (splicing) of Brieskorn homology spheres (Bh-spheres) \cite{SavB2} (see also Appendix).
Suppose that the set of nodes $\mathcal{N}_p$ of the graph $\Gamma_p$ is non-empty. Considering the graph $\Gamma_p$ as a one-dimensional simplicial complex, we take the complement $\Gamma_p - \mathcal{N}_p$. This complement is the disjoint union of straight line segments which are the maximal chains of $\Gamma_p$. If the chain is open at both extremities we call it internal maximal chain (see the chain between the nodes $N^I$ and $N^J$ in Fig \ref{f1}). An internal chain corresponds to a Seifert fibered thick  torus (homeomorphic to $T^2\times [0,1]$) in graph manifold $M(\Gamma_p)$. If the chain is half open we call it a terminal maximal chain, it corresponds to a Seifert fibered solid torus (homeomorphic to $D^2\times S^1$) in $M(\Gamma_p)$ (see example in Fig. \ref{f2}).

Now recall that each edge $\sigma$ of $\Gamma_p$ corresponds to the embedded torus $T^2_{\sigma}\subset M(\Gamma_p)$ and the collection of all these tori cuts the graph manifold $M(\Gamma_p)$ into disjoint union of circle bundles over $n$ times punctured sphere 
$$S^2 \setminus \bigcup^n_{l=1} p_l,~~~ 1\leq n\leq 3,$$ 
In general, the bundles are over compact surfaces of genus $g$ with some boundary components, see \cite{{Hirz}, {NeumLect}}. Such a collection of tori $\mathcal{T}_W$ is called a graph structure on $M(\Gamma_p)$ by Waldhausen \cite{Wald}. We want to define (following to \cite{PaPoPa}) a subcollection $\mathcal{T}_{JSJ} \subset \mathcal{T}_W$ of the Waldhausen graph structure, which is known as Jaco-Shalen-Johannson (JSJ) graph structure and to specify the corresponding JSJ-decomposition of graph manifold $M(\Gamma_p)$ on the set of Seifert fibered pieces $M_{JSJ}(N^I)$. Let us denote $\mathcal{C}(p)$ the set of maximal chains in the graph $\Gamma_p$. This set can be written as a disjoint union 
$$
\mathcal{C}(p)=\mathcal{C}_i(p)\bigsqcup\mathcal{C}_t(p),$$
where $\mathcal{C}_i(p)$ denotes the set of interior chains and $\mathcal{C}_t(p)$ is the set of terminal chains. The edges of $\Gamma_p$ contained in a chain $C\in \mathcal{C}(p)$ correspond to a set of parallel tori in $M(\Gamma_p)$. Choose one torus $T_C^2$ among them and define 
\be\label{2.11}\mathcal{T}_{JSJ}=\bigsqcup_{C\in\mathcal{C}_i(p)}T_C^2 \ee
This set of tori performs the well known JSJ-decomposition of the graph manifold $M(\Gamma_p)$ \cite{PaPoPa}. 

By construction, each piece $M_{JSJ}(N^I)$ (which we shall denote as $M_{JSJ}^I$ for brevity) of JSJ-decomposition that corresponds to the node $N^I$ contains a unique piece $M_W(N^I)$ (which we shall denote as $M_W^I$) of Waldhausen decomposition associated with the same node $N^I$. One can extend in a unique way up to isotopy the natural Seifert structure without exceptional fibers on $M_W^I$ to a Seifert fibration on $M_{JSJ}^I$ with exceptional fibers. Thus in these terms the JSJ-decomposition of the manifold $M(\Gamma_p)$ is defined completely by
$$M(\Gamma_p)=\bigcup_{I=1}^R\bar{M}_{JSJ}^I,$$
where $R$ is the number of nodes in $\Gamma_p$ and the bar over $M$ means the closure of the open piece $M_{JSJ}^I$. 

\begin{figure}[h]

\begin{center}

\setlength{\unitlength}{1pt}
\begin{picture}(300,50)
\put(10,-0.5){\makebox(.5,.5){$\bullet$}}
\put(10,0){\line(1,0){25}}
\put(25,-7){\makebox(.5,.5){$a_3^1$}}

\put(40,38){\makebox(.5,.5){$\bullet$}}
\put(40,5){\line(0,1){35}}
\put(50,15){\makebox(.5,.5){$a_1^1$}}

\put(40,0){\circle{10}}

\put(45,0){\line(1,0){25}}
\put(55,-7){\makebox(.5,.5){$a_2^1$}}

\multiput(80,-0.5)(10,0){3}{.}

\put(110,0){\line(1,0){25}}
\put(125,-7){\makebox(.5,.5){$a_3^I$}}

\put(140,5){\line(0,1){35}}
\put(150,15){\makebox(.5,.5){$a_1^I$}}
\put(140,38){\makebox(.5,.5){$\bullet$}}

\put(140,0){\circle{10}}

\put(145,0){\line(1,0){35}}
\put(150,-7){\makebox(.5,.5){$a_2^I$}}
\put(170,-7){\makebox(.5,.5){$a_3^{I+1}$}}

\put(185,5){\line(0,1){35}}
\put(198,15){\makebox(.5,.5){$a_1^{I+1}$}}
\put(185,38){\makebox(.5,.5){$\bullet$}}

\put(185,0){\circle{10}}

\put(190,0){\line(1,0){25}}
\put(200,-7){\makebox(.5,.5){$a_2^{I+1}$}}

\multiput(225,-0.5)(10,0){3}{.}

\put(255,0){\line(1,0){25}}
\put(270,-7){\makebox(.5,.5){$a_3^R$}}

\put(285,5){\line(0,1){35}}
\put(295,15){\makebox(.5,.5){$a_1^R$}}
\put(285,38){\makebox(.5,.5){$\bullet$}}

\put(285,0){\circle{10}}

\put(290,0){\line(1,0){25}}
\put(295,-7){\makebox(.5,.5){$a_2^R$}}
\put(315,-0.5){\makebox(.5,.5){$\bullet$}}

\end{picture}\\

\vspace*{1.cm}
\caption{A plumbing (splicing) diagram $\Delta_p$. \label{f3}}
\end{center}
\end{figure}
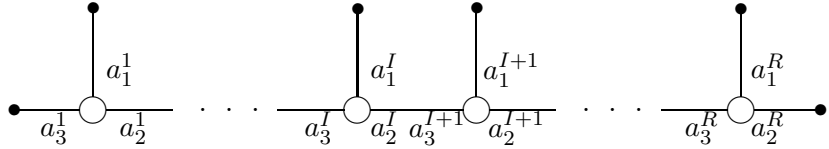

Note that there exists an uncertainty in the choice of the torus $T_C^2$ for each internal chain which appear in the JSJ-structure (\ref{2.11}). We can remove  this uncertainty in following way. Let us perform the {\it maximal} extension of the natural Seifert fibration from each $M_W^I$ and denote the obtained Seifert fibered piece of $M(\Gamma_p)$ by $\hat{M}^I$. It is clear that 
$\hat{M}^I\cap\hat{M}^J\neq \emptyset$ if and only if there exists a chain $C_{IJ}$ joining the nodes $N^I$ and $N^J$. 
If we start with plumbing of $R$ Bh-spheres  $\{\Sigma(a_1^I,
a_2^I,a_3^I)| I=1,\dots, R\}$, the resulting graph three-manifold will be integer homology sphere \cite{{SavB2}, {NeumLect}} ($\mathbb Z$-homology sphere), which in general case does not have the global Seifert fibration. But we can construct the JSJ-covering 
\be\label{2.12}\mathcal{M}:=\{\hat{M}^I|I=1,\dots, R\},\ee
 such as each $\hat{M}^I$ is a Seifert fibered space and it is {\it maximal} in the sense described above. Suppose that we perform the plumbing operation according to the plumbing (splicing) diagram $\Delta_p$, shown on the Fig. \ref{f3}. Thus our plumbing diagrams will always have the pairwise coprime weights around each node and correspond to $\mathbb Z$-homology spheres \cite{EisNeum}.

We suppose also that the order of the Seifert invariants $a_i^I$ is the following one: $a_1^1<a_3^1 <a_2^1$ and $a_1^I<a_2^I<a_3^I$ for $I\geq 2$. In this case the integer linking matrix \footnote{We remind the definition of integer linking matrix in the Remark 2.5.} (intersection form) $Q(\Gamma_p)$ which corresponds to the plumbing graph $\Gamma_p$ is definite (positive or negative) if and only if
\be\label{2.14} a_2^I a_3^{I+1}> a_1^I a_3^I a_1^{I+1}a_2^{I+1},\ee 
for $I=1,\dots,R-1$ (see the Proof of this statement in \cite{EisNeum}).
W. Neumann showed in \cite{Neum77}  that the same condition ensures that the reduced plumbing matrix $K^{IJ}_{\rm reduced}$ is definite.

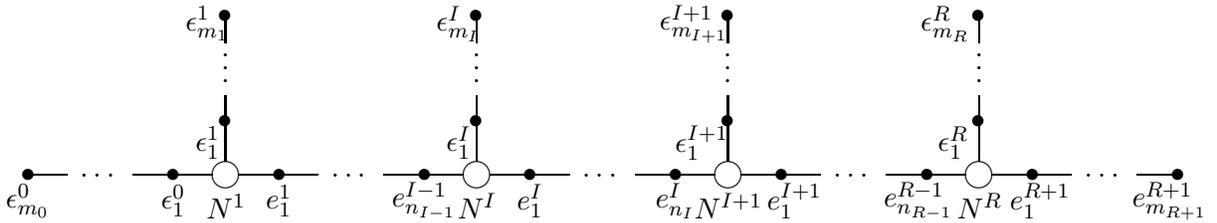
\begin{figure}[h]
\begin{center}

\setlength{\unitlength}{1pt}
\begin{picture}(440,50)
\put(0,-0.5){\makebox(.5,.5){$\bullet$}}
\put(0,0){\line(1,0){15}}
\put(0,-10){\makebox(.5,.5){$\epsilon^0_{m_0}$}}

\multiput(20,-0.5)(5,0){3}{.}

\put(40,0){\line(1,0){15}}
\put(55,-0.5){\makebox(.5,.5){$\bullet$}}
\put(55,0){\line(1,0){15}}
\put(55,-10){\makebox(.5,.5){$\epsilon^0_1$}}

\put(75,0){\circle{10}}
\put(75,-12){\makebox(.5,.5){$N^1$}}

\put(74.5,20){\makebox(.5,.5){$\bullet$}}
\put(75,5){\line(0,1){15}}
\put(68,12){\makebox(.5,.5){$\epsilon^1_1$}}
\put(75,20){\line(0,1){10}}
\multiput(73.5,35)(0,5){3}{.}
\put(75,50){\line(0,1){10}}
\put(68,57){\makebox(.5,.5){$\epsilon^1_{m_1}$}}
\put(74.5,60){\makebox(.5,.5){$\bullet$}}

\put(80,0){\line(1,0){15}}
\put(95,-0.5){\makebox(.5,.5){$\bullet$}}
\put(95,0){\line(1,0){15}}
\put(95,-10){\makebox(.5,.5){$e^1_1$}}

\multiput(115,-0.5)(5,0){3}{.}

\put(135,0){\line(1,0){15}}
\put(150,-0.5){\makebox(.5,.5){$\bullet$}}
\put(150,0){\line(1,0){15}}
\put(150,-10){\makebox(.5,.5){$e^{I-1}_{n_{I-1}}$}}

\put(170,0){\circle{10}}
\put(170,-12){\makebox(.5,.5){$N^I$}}

\put(169.5,20){\makebox(.5,.5){$\bullet$}}
\put(170,5){\line(0,1){15}}
\put(163,12){\makebox(.5,.5){$\epsilon^I_1$}}
\put(170,20){\line(0,1){10}}
\multiput(168.5,35)(0,5){3}{.}
\put(170,50){\line(0,1){10}}
\put(163,57){\makebox(.5,.5){$\epsilon^I_{m_I}$}}
\put(169.5,60){\makebox(.5,.5){$\bullet$}}

\put(175,0){\line(1,0){15}}
\put(190,-0.5){\makebox(.5,.5){$\bullet$}}
\put(190,0){\line(1,0){15}}
\put(190,-10){\makebox(.5,.5){$e^I_1$}}

\multiput(210,-0.5)(5,0){3}{.}

\put(230,0){\line(1,0){15}}
\put(245,-0.5){\makebox(.5,.5){$\bullet$}}
\put(245,0){\line(1,0){15}}
\put(245,-10){\makebox(.5,.5){$e^{I}_{n_I}$}}

\put(265,0){\circle{10}}
\put(265,-12){\makebox(.5,.5){$N^{I+1}$}}

\put(264.5,20){\makebox(.5,.5){$\bullet$}}
\put(265,5){\line(0,1){15}}
\put(255,12){\makebox(.5,.5){$\epsilon^{I+1}_1$}}
\put(265,20){\line(0,1){10}}
\multiput(263.5,35)(0,5){3}{.}
\put(265,50){\line(0,1){10}}
\put(252,57){\makebox(.5,.5){$\epsilon^{I+1}_{m_{I+1}}$}}
\put(264.5,60){\makebox(.5,.5){$\bullet$}}

\put(270,0){\line(1,0){15}}
\put(285,-0.5){\makebox(.5,.5){$\bullet$}}
\put(285,0){\line(1,0){15}}
\put(290,-10){\makebox(.5,.5){$e^{I+1}_1$}}
\multiput(305,-0.5)(5,0){3}{.}

\put(325,0){\line(1,0){15}}
\put(340,-0.5){\makebox(.5,.5){$\bullet$}}
\put(340,0){\line(1,0){15}}
\put(337,-10){\makebox(.5,.5){$e^{R-1}_{n_{R-1}}$}}

\put(360,0){\circle{10}}
\put(360,-12){\makebox(.5,.5){$N^R$}}

\put(359.5,20){\makebox(.5,.5){$\bullet$}}
\put(360,5){\line(0,1){15}}
\put(350,12){\makebox(.5,.5){$\epsilon^R_1$}}
\put(360,20){\line(0,1){10}}
\multiput(358.5,35)(0,5){3}{.}
\put(360,50){\line(0,1){10}}
\put(347,57){\makebox(.5,.5){$\epsilon^R_{m_R}$}}
\put(359.5,60){\makebox(.5,.5){$\bullet$}}

\put(365,0){\line(1,0){15}}
\put(380,-0.5){\makebox(.5,.5){$\bullet$}}
\put(380,0){\line(1,0){15}}
\put(383,-10){\makebox(.5,.5){$e^{R+1}_1$}}

\multiput(400,-0.5)(5,0){3}{.}

\put(420,0){\line(1,0){15}}
\put(435,-0.5){\makebox(.5,.5){$\bullet$}}
\put(432,-10){\makebox(.5,.5){$e^{R+1}_{m_{R+1}}$}}

\end{picture}\\
\vspace*{1.cm}
\caption{The plumbing graph $\Gamma_p$ corresponding to plumbing diagram $\Delta_p$. \label{f4}}
\end{center}

\end{figure}

In order to construct a plumbing graph $\Gamma_p$ for a $\mathbb{Z}$-homology sphere, we follow to the procedure described in \cite{{EisNeum}, {Hirz}, {NeumLect}}. First of all we must calculate the characteristics of maximal chains. For terminal chains the integer Euler numbers $\epsilon_i^I$ are defined by the continued fraction:  
\be\label{2.15} -\frac{a^I}{ b^I}=[\epsilon_1^I,\dots,\epsilon_{m_I}^I],\ee
here $(a^I, b^I)$, $I=0,\dots,R+1$ are the Seifert invariants, numerated in the following way
\be\label{2.16} a^0=a^1_3,~ b^0=b^1_3;~ a^J=a^J_1,~ b^J=b^J_1,~ J=1,\dots,R;~ a^{R+1}=a^R_2,~ b^{R+1}=b^R_2,\ee
in order to eliminate the inferior indexes. It is appropriate to remind that a continued fraction is defined as
$$
-p/q=[n_1,\dots,n_s]=n_1-\frac{1}{n_2-\displaystyle\frac{1}{\cdots-
\displaystyle\frac{1}{n_s}}},
$$
For internal chains the integer Euler numbers $e_i^I$ are defined by
\be\label{2.17} -\frac{p^I}{q^I}=[e_1^I,\dots,e_{n_I}^I],\ee
where the Seifert (orbital) invariants $(p^I, q^I)$, $I=1,\dots,R-1$ characterize the {\it thick tori} $TT(p^I, q^I)\cong T^2\times [0,1]$, which are created by the plumbing operations performed between the nodes $N^I$ and $N^{I+1}$, see \cite{{SavArt}, {NeumLect}}.
These invariants identify also the extra lens spaces $L(p^I, q^I)$ which arise as a component of the boundary of four-dimensional plumbed V-cobordism (corresponding to the graph $\Gamma_p$) constructed in \cite{SavArt}. In the same paper N. Saveliev gave the formulas for 
$p^I$ and $q^I$: 
\be\label{2.18} p^I= a_2^I a_3^{I+1}- a_1^I a_3^I a_1^{I+1}a_2^{I+1},~~~~
q^I= b_2^I a_3^{I+1} + a_1^{I+1}a_2^{I+1}(b_1^I a_3^I + a_1^I b_3^I) \ee
for the ordering fixed by the plumbing diagram in Fig. \ref{f3}.  
The plumbing graph $\Gamma_p$, corresponding to the plumbing (splicing) diagram from the Fig. \ref{f3}, is shown in Fig. \ref{f4}.

~~~~~~\\

The Euler numbers $\epsilon_i^I$ and $e_i^I$ decorate the vertices with valence 1 and 2. The nodes are marked by $N^I$ with $I=1,\dots,R$. We shall used the innormalized
Seifert invariants, thus the Euler number associated with each node is zero (see \cite{SavB1} and Appendix for more details).
From this representation of the plumbing graph it is clear that for $I=2,\dots,R-1$ the set $\hat{M}^I$ of JSJ-covering $\mathcal{M}$ has the form
\be\label{2.19}\hat{M}^I=\bar{M}^I_W\cup TT(p^{I-1},{q*}^{I-1})\cup TT(p^I,q^I)\cup ST(a^I,b^I),  \ee
where $ST(a^I,b^I)$ is a Seifert fibered solid torus with Seifert invariants $(a^I,b^I)$ and 
\be\label{2.20} -\frac{p^I}{{q*}^I}=[e_{n_I}^I,\dots,e_1^I],\ee
For the cases $I=1$ and $I=R$ the formulas are different from (\ref{2.19}):
\be\label{2.21}\hat{M}^1=\bar{M}^1_W\cup TT(p^1,q^1)\cup ST(a^0,b^0)\cup ST(a^1,b^1),  \ee
\be\label{2.22}\hat{M}^R=\bar{M}^R_W\cup TT(p^{R-1},{q*}^{R-1})\cup ST(a^R,b^R)\cup ST(a^{R+1},b^{R+1}), \ee
see the notations in (\ref{2.16}). Moreover
\be\label{2.23}\hat{M}^I\cap \hat{M}^{I+1}=TT(p^I,q^I)\cong^{*} TT(p^I,{q*}^I)~~ I=2,\dots,R-1 .\ee
Here the symbol $\cong^{*}$ indicates that $TT(p^I,q^I)$ and $TT(p^I,{q*}^I)$ are homeomorphic, but their Seifert structures are characterized by different integer Euler numbers defined by
$$-\frac{p^I}{q^I}=[e_1^I,\dots,e_{n_I}^I]~~~ {\rm and}~~~ -\frac{p^I}{{q*}^I}=[e_{n_I}^I,\dots,e_1^I]$$ respectively.
Thereby the thick torus between the nodes $N^I$
and $N^{I+1}$ has two Seifert fibrations: the first is the extension of the natural Seifert fibration defined on the piece $M_W^I$ and the second one is obtained as extension from the piece $M_W^{I+1}$. These Seifert fibrations are connected by the matrix \cite{{NeumLect}, {SavArt}}:   
\be \label{2.24} S^I=\left(
\begin{array}{ll}
-q^I & {p*}^I\\
~~p^I & {q*}^I
\end{array}
\right)=
\left(
\begin{array}{ll}
-1 & 0\\
-e_1 & 1\end{array}
\right)
\left(
\begin{array}{ll}
0 & 1\\
1 & 0\end{array}
\right)
\left(
\begin{array}{ll}
-1 & 0\\
-e_2 & 1\end{array}
\right)\dots \left(
\begin{array}{ll}
0 & 1\\
1 & 0\end{array}
\right)\left(
\begin{array}{ll}
-1 & 0\\
-e_n & 1\end{array}
\right)\ee
in the following sense.
Recall that edges of $\Gamma_p$ contained in a chain $C_{I,I+1}$ (between the nodes $N^I$ and $N^{I+1}$) correspond to the set parallel tori in $M(\Gamma_p)$. On any of these tori there exist two bases formed by the section lines and the fibers pertain to the Seifert fibrations extended from $M^I_W$ and $M^{I+1}_W$, which we denote as the pair of columns
\be \label{2.25}
\left(
\begin{array}{l}
s^I_2\\
f_I
\end{array}
\right)~~~~{\rm and}~~~~
\left(
\begin{array}{l}
s^{I+1}_3\\
f_{I+1}
\end{array}
\right).\ee
Subindices 2 and 3 manifest that $\Sigma(a_1^I, a_2^I, a_3^I)$ and $\Sigma(a_1^{I+1}, a_2^{I+1}, a_3^{I+1})$ are plumbed together along the singular fibers with Seifert invariants $a_2^I$ and $a_3^{I+1}$ (see Fig.~\ref{f3}). Then the transformation between these section-fiber bases is described by 
\be \label{2.26}
\left(
\begin{array}{l}
s^I_2\\
f_I
\end{array}
\right)=\left(
\begin{array}{ll}
-q^I & {p*}^I\\
~~p^I & {q*}^I
\end{array}
\right)\left(
\begin{array}{l}
s^{I+1}_3\\
f_{I+1}
\end{array}
\right),\ee where ${p*}^I$ is defined from $\det S^I =-(q^I{q*}^I+p^I{p*}^I)=-1$.

Now we introduce the one-form bases
\be \label{2.27} (\sigma^2_I, \kappa^I)~~~~\rm{and}~~~~(\sigma^3_{I+1}, \kappa^{I+1}),  \ee 
duals to the bases (\ref{2.25}) in the following sense:
\be \label{2.28} \int_{s_2^I}\sigma^2_I=\int_{s_3^{I+1}} \sigma^3_{I+1}=\int_{f_I} \kappa^I=1;  \ee 
\be \label{2.29} \int_{f_I}\sigma^2_I=\int_{f_{I+1}} \sigma^3_{I+1}=
\int_{s_2^I} \kappa^I=\int_{s_3^{I+1}} \kappa^{I+1}=0,  \ee 
where the integrals are calculated over any such section line or fiber as, for example, in \cite{BeasleyWitt}. Thus we obtain the corresponding transformations between the the dual one-forms:
\be \label{2.30} \sigma^2_I=-{q*}^I \sigma^3_{I+1}+ p^I\kappa^{I+1};~~~~~~  
\kappa^I={p*}^I\sigma^3_{I+1}+q^I\kappa^{I+1}. \ee

We have to clarify the question about the supports of one-forms $\kappa^I$ and $\sigma^i_I$, ($i=1,2,3$). From the definition (\ref{2.12}) of the JSJ-covering $\mathcal{M}$ we conclude that the one-form $\kappa^I$, which is connected with the fibers, has the support $\hat{M}^I$, ($I=1,\dots R$). Note that the Seifert fibered manifold $\hat{M}_I$ can be identified with the total space of principal $U(1)$ V-bundle \cite{GriTan} associated with the line V-bundle over the orbifold $\hat{\Sigma}^I:=\hat{M}_I/S^1$. Thus, following to \cite{BeasleyWitt} we regard the one-form $\kappa^I$ as a connection on the total space $\hat{M}_I$. From the decomposition (\ref{2.19}) the supports of the one-forms $\sigma^2_I$ and $\sigma^3_{I+1}$ are $M^2_I=\bar{M}^I_W\cup TT(p^I,q^I)$ and $M^3_{I+1}=\bar{M}^{I+1}_W\cup TT(p^I,{q*}^I)$, respectively for $I=1,\dots, R-1$.
The one-form $\sigma^2_R$ has the support $M^2_R=\bar{M}^R_W\cup ST(a^{(R+1)},b^{(R+1)})$ and $\sigma^3_1$ has the support $M^3_1=\bar{M}^1_W\cup ST(a^0,b^0)$.
Finally, the supports of the one-forms $\sigma^1_I$ are $M^1_I=\bar{M}^I_W\cup ST(a^I,b^I)$ for $1\leq I\leq R$ (see the notations of $a^I$ and $b^I$ the formulas in (\ref{2.16}).\\

\noindent{\bf Remark 2.2.}
Hereafter we suppose that the Seifert fibration on each $\bar{M}^I_W$ with $I=1,\dots, R$ is trivial. This case corresponds to the assumption that unnormalized Seifert invariants are used everywhere in this paper (see  \cite{{EisNeum}, {SavB1}} and Appendix).\\

We suppose, further, that the forms $\sigma$ and $\kappa$ are dual with respect to the bilinear pairing defined as
\be \label{2.31} <\sigma^2_I, \kappa^J> := \int_{\hat{M}^I\cap\hat{M}^J}\sigma^2_I\wedge d\kappa^J = \int_{M^2_I\cap\hat{M}^J}\sigma^2_I\wedge d\kappa^J = \delta_I^J; \ee
\be \label{2.32} <\sigma^3_{I+1}, \kappa^J> :=\int_{\hat{M}^{I+1}\cap\hat{M}^J}\sigma^3_{I+1}\wedge d\kappa^J = \int_{M^3_{I+1}\cap\hat{M}^J}\sigma^3_{I+1}\wedge d\kappa^J = \delta_{I+1}^J, \ee
For the graph manifold with plumbing graph shown in Fig. \ref{f4} the only nonempty intersections are:
$$M^2_I\cap\hat{M}^I=\bar{M}^I_W\cup TT(p^I, q^I);~~~
M^2_I\cap\hat{M}^{I+1}=TT(p^I, q^I)$$
$$M^3_{I+1}\cap\hat{M}^{I+1}=\bar{M}^{I+1}_W\cup TT(p^I, {q*}^I);~~~
M^3_{I+1}\cap\hat{M}^I=TT(p^I, {q*}^I)$$
Note that, since the submanifolds $\bar{M}^I_W$, for $I=1,\dots,R$, have (according to the supposition in Remark 2.2) the trivial Seifert fibrations, their contributions to the integrals (\ref{2.31}) and (\ref{2.32}) are zero.  

Also we shall used the integrals 
\be\label{2.33} \Lambda^{I,I+1}=\int_{TT(p^I, q^I)}\kappa^I\wedge d\kappa^{I+1}=\int_{L(p^I, q^I)}\kappa^I\wedge d\kappa^{I+1},~~~I=1,\dots, R-1;
\ee
\be\label{2.34} \Lambda^{I,I}=\int_{TT(p^I, q^I)}\kappa^I\wedge d\kappa^{I}=\int_{L(p^I, q^I)}\kappa^I\wedge d\kappa^{I},~~~I=1,\dots, R-1;
\ee
\be\label{2.35} \Lambda^{I+1,I+1}=\int_{TT(p^I, {q*}^I)}\kappa^{I+1}\wedge d\kappa^{I+1}=\int_{L(p^I, {q*}^I)}\kappa^{I+1}\wedge d\kappa^{I+1},~~~I=1,\dots, R-1,
\ee
which define the linking (intersection) numbers of the fiber structures $\kappa^I$ and $\kappa^{I+1}$ defined on thick torus $TT(p^I, q^I)\cong^{*} TT(p^I, {q*}^I)$ and on the corresponding lens space $L(p^I, q^I)\cong^{*} L(p^I, {q*}^I)$ (the homeomorphism of these lens spaces is a consequence of $q^I{q*}^I+p^I{p*}^I=1$).
The integrals over $TT(p^I, q^I)$ and $L(p^I, q^I)$ as well as over $TT(p^I, {q*}^I)$ and $L(p^I, {q*}^I)$ are equal since the lens space can be represented as 
$$ L(p^I, q^I)=ST_{\rm{in}}\cup_J TT(p^I, q^I)\cup_JST_{\rm{fin}}$$
\be\label{2.36} L(p^I, {q*}^I)=ST_{\rm{in}}\cup_J TT(p^I, {q*}^I)\cup_JST_{\rm{fin}},\ee
where $ST_{\rm{in}}$ and $ST_{\rm{fin}}$ are solid tori with trivial Seifert fibrations plumbed to the thick torus $TT(p^I, q^I)$ and $TT(p^I, {q*}^I)$ respectively; thus their contributions to the integrals are null. Consequently, the integrations  in (\ref{2.31}) - (\ref{2.35}) effectively are performed over thick torus $TT(p^I, q^I)$ or $TT(p^I,{q*}^I)$, where the both fibrations $(\kappa^I, \sigma_I^2)$ and $(\kappa^{I+1}, \sigma_{I+1}^3)$ are defined.\\

\noindent{\bf Remark 2.3.} Since, by definition of the Seifert fibration on  Bh-spheres, there exists smooth continuations of the sections $s^I_i$, $i=1,2,3$ to the section of the trivial Seifert fibration on $M^I_W$ \cite{EisNeum}, then the dual one-forms $\sigma_I^i, i=1,2,3$ unite into the one-form $\sigma_I$ which is defined over the whole orbifold $\hat{\Sigma}^I$.\\

We can obtain the rational linking matrix for $TT(p^I,q^I)\cong^{*} TT(p^I,{q*}^I)$, or equivalently for $L(p^I,q^I)\cong^{*} L(p^I,{q*}^I)$ by means of multiplication of the equations (\ref{2.30}) by $d\kappa^I$ and $d\kappa^{I+1}$ and integration over $TT(p^I,q^I)\cong^{*} TT(p^I,{q*}^I)$ (equivalently over $L(p^I,q^I)\cong^{*} L(p^I,{q*}^I)$). 
Applying the duality conditions (\ref{2.31}) and (\ref{2.32}) we obtain:    
\be\label{2.37} \Lambda^{I,I+1}=\Lambda^{I,I+1}= \frac{1}{p^I}~~~ 
\Lambda^{I,I}= \frac{q^I}{p^I};~~~ \Lambda^{I+1,I+1}=\frac{{q*}^I}{p^I}.   \ee  
The rational numbers $\Lambda^{I,I}$ and $\Lambda^{I+1,I+1}$ are also known as Chern classes of the line V-bundles associated to the Seifert fibrations andoved with the $U(1)-$invariant connection forms $\kappa^{I}$ and $\kappa^{I+1}$ on the lens spaces $L(p^I, q^I)$ and $L(p^I, {q*}^I)$ respectively \cite{BeasleyWitt}.  This result for the lens spaces was obtained by other method in \cite{GriTan} and it gives the possibility to calculate rational linking matrices for graph manifolds corresponding to the tree graph $\Gamma_p$, shown in Fig. \ref{f4}.

\subsection{Rational linking matrices for graph manifolds} \label {s2.2}
Let us introduce the basic notion of this paper, namely, rational linking matrix for the graph manifold $M^3_{+}=-M(\Gamma_p)$ (see Fig. \ref{f4}): 
\be\label{2.38} K^{IJ}=\int_{M^3_{+}}\kappa^I\wedge d\kappa^{J} =-\int_{M(\Gamma_p)}\kappa^I\wedge d\kappa^{J}.\ee
We integrate here over the three dimensional graph manifold $M^3_{+}$ possessing the positive definite linking matrix. This manifold has the opposite orientation with respect to the graph manifold $M(\Gamma_p)$ obtained directly by plumbing of Bh-spheres, which are defined as links of singularities (see Appendix and equation (\ref{4.1})).
This construction of the graph manifold $M(\Gamma_p)$ gives the possibility to represent it also as a link of singularity that guarantees its rational linking matrix to be negative definite when the condition (\ref{2.14}) is fulfilled for all $I$ \cite{EisNeum}. From the tree structure of the graph $\Gamma_p$, and from the first equation in (\ref{2.37}) we immediately obtain, that for $I\neq J$ the nonzero elements are only
\be\label{2.39} K^{I,I+1}=K^{I+1,I}=-\int_{\hat{M}^I\cap\hat{M}^{I+1}}\kappa^I\wedge d\kappa^{I+1}=-\int_{TT(p^I, q^I)}\kappa^I\wedge d\kappa^{I+1}= -\frac{1}{p^I},\ee
for $1\leq I\leq R-1.$

If $I=J=2,\dots, R-1$, we have 
$$K^{II}=-\int_{\hat{M}^I}\kappa^I\wedge d\kappa^{I}=$$
\be\label{2.40}-\int_{TT(p^{I-1}, {q*}^{I-1})}\kappa^I\wedge d\kappa^{I}- \int_{TT(p^I, q^I)}\kappa^I\wedge d\kappa^{I}- \int_{ST(a^I, b^I)}\kappa^I\wedge d\kappa^{I}.  \ee
Here we use the decomposition (\ref{2.19}) of the piece $\hat{M}^I$, and that the integral over trivial Seifert fibration $\bar{M}_W^I$ is zero. Then according to the two last equations in (\ref{2.37}) we obtain the matrix element
\be\label{2.41} K^{II}= -\left(\frac{{q*}^{I-1}}{p^{I-1}}+\frac{q^I}{p^I}+\frac{b^I}{a^I}\right), \ee 
also known as the Chern class of line $V$-bundle associated with the Seifert fibration of $\hat{M}^I$.
For $I=1$ and $I=R$ the matrix elements are
$$ K^{11}=-\int_{\hat{M}^1}\kappa^I\wedge d\kappa^{I}=$$
\be\label{2.42}-\int_{ST(a^0, b^0)}\kappa^1\wedge d\kappa^{1}- \int_{TT(p^1, q^1)}\kappa^1\wedge d\kappa^{1}- \int_{ST(a^1, b^1)}\kappa^I\wedge d\kappa^{I};\ee
$$K^{RR}=-\int_{\hat{M}^R}\kappa^R\wedge d\kappa^{R}=$$
\be\label{2.43}-\int_{TT(p^{R-1}, {q*}^{R-1})}\kappa^R\wedge d\kappa^{R}- \int_{ST(a^R, b^R)}\kappa^R\wedge d\kappa^{R}- \int_{ST(a^{R+1}, b^{R+1})}\kappa^R\wedge d\kappa^{R};\ee
\be\label{2.44}K^{11}=-\left(\frac{b^0}{a^0}+\frac{q^1}{p^1}+\frac{b^1}{a^1}\right); \ee
\be\label{2.45}K^{RR}=-\left(\frac{{q*}^{R-1}}{p^{R-1}}+\frac{b^R}{a^R}+\frac{b^{R+1}}{a^{R+1}}\right).  \ee
Here we have used the decompositions (\ref{2.21}) and (\ref{2.22}) as well as the notations (\ref{2.16}).\\

\noindent{\bf Remark 2.4.} The manifold $M(\Gamma_p)$ described by a graph $\Gamma_p$ of type, shown in Fig. \ref{f4}, has a JSJ-covering $\mathcal{M}$ (see formula (\ref{2.12})) which consist of $R$ pieces $\hat{M}^I$ with defined Seifert fibrations characterized by one-forms $\kappa^I$. But $M(\Gamma_p)$ does not have, in general, a global Seifert fibration. In this sense the formulas (\ref{2.41}), (\ref{2.44}) and (\ref{2.45}), determining the rational linking matrix, generalize the formula (3.22) from the paper of C. Beasley and E. Witten \cite{BeasleyWitt}:
$$ c_1(\hat{\mathcal{L}})=\int_{M}\kappa\wedge d\kappa=n_0+\sum_{j=1}^N \frac{\beta_i}{\alpha_i}$$
for the Chern class of the line V-bundle $\hat{\mathcal{L}}$ over the two-dimensional orbifold $\hat{\Sigma}$. This line V-bundle $\hat{\mathcal{L}}$ is associated with the $U(1)$ V-bundle over $\hat{\Sigma}$ with the total space $M$, {\it i.e.} the manifold $M$ has a global Seifert fibration, characterized by the normalized Seifert invariants $(\alpha_i, \beta_i)$, $i=1,\dots N$ and the degree of the line V-bundle $n_0$. It is always possible to reduce the degree $n_0$ to zero by using the unnormalized Seifert invariants \cite{EisNeum}, (see also Appendix) as we are doing in this paper.
(The integer Euler number $b$ introduced in Appendix is equal to $-n_0$.)
In other words, the rational linking matrix (\ref{2.38}), which is the topological invariant of the graph manifold $M(\Gamma_p)$, is the generalization of the unique Chern class that characterize the Seifert fibered manifold $M$. \\

\noindent{\bf Remark 2.5.} Remind that for the general case of a tree graph manifold $M(\Gamma_p)$, the integer linking matrix is defined as follows \cite{{Hirz}, {Orlik}}
$$Q^{AB}(\Gamma_p)=\left\{ 
\begin{array}{cl}
e_{A}, & \text{if }A=B\text{;} \\ 
-1, & \text{if }A\neq B\text{ and }v_{A}\text{ is connected to }v_{B}\text{
by an edge;} \\ 
0, & \text{otherwise,}
\end{array}
\right.$$
with integer Euler numbers $e_A$ corresponding to each vertex $v_A$ (see Fig. 4).\\

The I-th fragment of the integer linking matrix which corresponds to the I-th piece of the graph $\Gamma_p$ shown in Fig. \ref{f4} is represented as

$$Q^{AB}(\Gamma_p)=$$
\be \label{2.46}
\left[ 
\begin{array}{ccccccccccccccccccc}
\ddots  &  &  &  &  &  &  &  &  &  &  &  &  &  &  &  &  &  & \\ 
& \epsilon _{1}^{I-1} & -1 &  &  &  &  &  &  &  &  &  &  &  &  &  &  &  & 
\\ 
& -1 & 0^{I-1} & -1 &  &  &  &  &  &  &  &  &  &  &  &  &  &  & \\ 
&  & -1 & e_{1}^{I-1} & -1 &  &  &  &  &  &  &  &  &  &  &  &  &  & \\ 
&  &  & -1 & \ddots  &  &  &  &  &  &  &  &  &  &  &  &  &  & \\ 
&  &  &  &  & \ddots  & -1 &  &  &  &  &  &  &  &  &  &  &  & \\ 
&  &  &  &  & -1 & e_{n_{I-1}}^{I-1} & 0 & 0 & \cdots  & 0 & -1 &  &  &  & 
&  &  & \\ 
&  &  &  &  &  & 0 & \epsilon _{m_{I}}^{I} & -1 &  &  & 0 &  &  &  &  & 
&  & \\ 
&  &  &  &  &  & 0 & -1 & \ddots  &  &  & \vdots  &  &  &  &  &  &  & \\ 
&  &  &  &  &  & \vdots  &  &  & \ddots  & -1 & 0 &  &  &  &  &  &  & \\ 
&  &  &  &  &  & 0 &  &  & -1 & \epsilon _{1}^{I} & -1 &  &  &  &  &  & & 
\\ 
&  &  &  &  &  & -1 & 0 & \cdots  & 0 & -1 & 0^{I} & -1 &  &  &  &  &  & \\ 
&  &  &  &  &  &  &  &  &  &  & -1 & e_{1}^{I} & -1 &  &  &  &  & \\ 
&  &  &  &  &  &  &  &  &  &  &  & -1 & \ddots  &  &  &  &  & \\ 
&  &  &  &  &  &  &  &  &  &  &  &  &  & \ddots  & -1 &  &  & \\ 
&  &  &  &  &  &  &  &  &  &  &  &  &  & -1 & e_{n_{I}}^{I} &  &  & \\ 
&  &  &  &  &  &  &  &  &  &  &  &  &  &  &  & \epsilon _{m_{I+1}}^{I+1} & -1 &\\ 
&  &  &  &  &  &  &  &  &  &  &  &  &  &  &  &  -1  & 0^{I+1} & -1\\
&  &  &  &  &  &  &  &  &  &  &  &  &  &  &  & & -1  & \ddots 
\end{array}
\right] \ee
Note that $0^I$ denotes an integer Euler number 0 corresponding to the node $N^I$
(we use unnormalized Seifert invariants, as it is explained in Appendix) and it is meant 0 on every empty place except -1 in the tridiagonal submatrices.

We can make an important assumption that the tridiagonal submatrices of type 
$$K^{ab}\left(\underline{e},I\right) =\left( 
\begin{array}{cccc}
e_{1}^{I} & -1 &  &  \\ 
-1 & \ddots  &  &  \\ 
&  & \ddots  & -1 \\ 
&  & -1 & e_{n_{I}}^{I}
\end{array}
\right) 
~~\text{ and }~~
K^{\alpha \beta }\left(\underline{\epsilon} ,I\right) =\left( 
\begin{array}{cccc}
\epsilon _{m_{I}}^{I} & -1 &  &  \\ 
-1 & \ddots  &  &  \\ 
&  & \ddots  & -1 \\ 
&  & -1 & \epsilon _{1}^{I}
\end{array}
\right)$$
describe the low-energy physics of hierarchical topological fluids with $n_I$ and $m_I$ levels respectively \cite{{WenZee}, {FujitaLi}}, which are put together according to the structure of the graph $\Gamma_p$. The topological fluids are also characterized by charge vectors $t^a(\underline{e},I)$ and $t^b(\underline{\epsilon},I)$, which together with $K$-matrices define the filling factors 
\be  \label{2.47}
\nu(\underline{e},I)=t^a(\underline{e},I)K_{ab}(\underline{e},I)t^b(\underline{e},I);~~~\nu(\underline{\epsilon},I)=t^{\alpha}(\underline{\epsilon},I)K_{\alpha \beta}(\underline{\epsilon},I)t^{\beta}(\underline{\epsilon},I), \ee
where $K_{ab}(\underline{e},I)$ and $K_{\alpha \beta}(\underline{\epsilon},I)$ are the matrices inverse to $K^{ab}(\underline{e},I)$ and $K^{\alpha \beta}(\underline{\epsilon},I)$ respectively.

If we make the choice $t^a(\underline{e},I)=\delta^a_1$, then
\be \label{2.48} \nu_1(\underline{e},I)=K_{11}(\underline{e},I)=\frac{\text{adj}K^{11}(\underline{e},I)}{\det K^{ab}(\underline{e},I)}=\frac{1}{[e^I_1,\dots,e^I_{n_I}]}=-\frac{q^I}{p^I}; \ee
choosing $t^a(\underline{e},I)=\delta^a_{n_I}$ we have
\be \label{2.49} \nu_n(\underline{e},I)=K_{n_I,n_I}(\underline{e},I)=\frac{\text{adj}K^{n_I,n_I}(\underline{e},I)}{\det K^{ab}(\underline{e},I)}=\frac{1}{[e^I_{n_I},\dots,e^I_1]}=-\frac{{q*}^I}{p^I}; \ee
and, finnaly for the choice $t^{\alpha}(\underline{\epsilon},I)=\delta^{\alpha}_1$ the filling factor is
\be \label{2.50} \nu_1(\underline{\epsilon},I)=K_{11}(\underline{\epsilon},I)=\frac{\text{adj}K^{11}(\underline{\epsilon},I)}{\det K^{\alpha \beta}(\underline{\epsilon},I)}=\frac{1}{[{\epsilon}^I_1,\dots,{\epsilon}^I_{m_I}]}=-\frac{b^I}{a^I}, \ee
see formulas (\ref{2.15}), (\ref{2.17}) and (\ref{2.20}).

Let us emphasize three important consequences which follow from our supposition.

The first one is that these filling factors coincide (up to sign) with the Chern classes of the line V-bundles associated to the Seifert fibrations with the $U(1)-$invariant connection forms on the lens spaces $L(p^I, q^I)$, $L(p^I, {q*}^I)$ and $L(a^I, b^I)$ respectively, as was mentioned after the equations (\ref{2.37}). 

The second fact is that the diagonal elements of rational linking matrix $K^{II}$ can be rewritten by means of filling factors of corresponding topological fluids:
\be \label{2.51}K^{II} = \nu_n(\underline{e},I-1)+\nu_1(\underline{e},I)+\nu_1(\underline{\epsilon},I),~~{\rm for}~~ I=2,\dots, R-1 \ee and
\be \label{2.52}K^{11} = \nu_1(\underline{\epsilon},0)+\nu_1(\underline{\epsilon},1)+\nu_1(\underline{e},1);~~~
K^{RR} = \nu_n(\underline{e},R-1)+\nu_1(\underline{\epsilon},R)+\nu_1(\underline{\epsilon},R+1),\ee
see the formulas (\ref{2.41}), (\ref{2.44}) and (\ref{2.45}). In the topological BF-model constructed in the Section 3, the rational linking matrix $K^{IJ}$ plays the role of gauge coupling constants matrix. In the Section 5 we shall show that, for the class of graph manifolds ($\mathbb Z$-homology spheres) constructing in this paper the following conditions are fulfilled
$$|K^{II}|\ll \min \{|\nu_n(\underline{e},I-1)|,|\nu_1(\underline{e},I)|,\nu_1(\underline{\epsilon},I)|\},~~{\rm for}~~ I=2,\dots, R-1; $$
\be \label{2.53}|K^{RR}|\ll \min \{|\nu_n(\underline{e},R-1)|,|\nu_1(\underline{\epsilon},R)|,|\nu_1(\underline{\epsilon},R+1)|\}.\ee 
We interpret this phenomenon as fine tuning for the gauge coupling constants described by $K^{IJ}$. This compensation effect occurs among the filling factors of the topological fluids corresponding to each node of the graph $\Gamma_p$.  In our model the fine tuning effect is universal for all coupling constants (not only for cosmological one) and is a result of special properties of topological invariants (self-linking numbers) characterizing the graph manifolds which form the extra dimensions in the Kaluza-Klein approach.

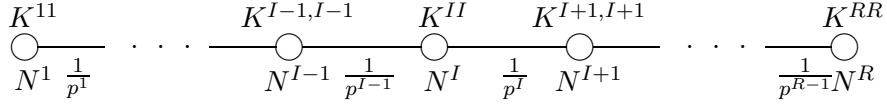
\begin{figure}[h]
\begin{center}

\setlength{\unitlength}{1pt}
\begin{picture}(310,50)

\put(-5,0){\circle{10}}
\put(-2,12){\makebox(.5,.5){$K^{11}$}}
\put(-2,-12){\makebox(.5,.5){$N^1$}}

\put(0,0){\line(1,0){25}}
\put(15,-12){\makebox(.5,.5){$\frac{1}{p^1}$}}

\multiput(35,-0.5)(10,0){3}{.}

\put(65,0){\line(1,0){25}}

\put(95,0){\circle{10}}
\put(98,12){\makebox(.5,.5){$K^{I-1,I-1}$}}
\put(98,-12){\makebox(.5,.5){$N^{I-1}$}}

\put(100,0){\line(1,0){45}}
\put(125,-12){\makebox(.5,.5){$\frac{1}{p^{I-1}}$}}

\put(150,0){\circle{10}}
\put(153,12){\makebox(.5,.5){$K^{II}$}}
\put(153,-12){\makebox(.5,.5){$N^{I}$}}

\put(155,0){\line(1,0){45}}
\put(180,-12){\makebox(.5,.5){$\frac{1}{p^I}$}}

\put(205,0){\circle{10}}
\put(208,12){\makebox(.5,.5){$K^{I+1,I+1}$}}
\put(208,-12){\makebox(.5,.5){$N^{I+1}$}}

\put(210,0){\line(1,0){25}}

\multiput(245,-0.5)(10,0){3}{.}

\put(275,0){\line(1,0){25}}
\put(290,-12){\makebox(.5,.5){$\frac{1}{p^{R-1}}$}}

\put(305,0){\circle{10}}
\put(308,12){\makebox(.5,.5){$K^{RR}$}}
\put(308,-12){\makebox(.5,.5){$N^{R}$}}

\end{picture}\\
\vspace*{1.cm}
\caption{A reduced plumbing graph $\Gamma^p_{\text{reduced}}$. \label{f5}}
\end{center}
\end{figure}

The third consequence is that the partially diagonalized matrix  $Q^{AB}_{\rm{part.diag}}(\Gamma_p)$ (equivalent to $Q^{AB}(\Gamma_p)$), obtained by the Gauss method in \cite{Neum77}, is represented as a direct sum  
\be \label{2.54}Q^{AB}_{\rm{part.diag}}(\Gamma_p) = K^{IJ}_{\rm{reduced}}\oplus D^{MN}.\ee 
Here the reduced plumbing matrix $K^{IJ}_{\rm{reduced}}$ coincides with the rational linking matrix $K^{IJ}$, since the elements of $K^{IJ}_{\rm{reduced}}$ have the same expressions (\ref{2.41}), (\ref{2.44}) and (\ref{2.45}) (see formula (*) on p. 366 in \cite{Neum77}).
The direct summand, $D^{MN}$, is the diagonal matrix such as
$$D^{MN}= \oplus_{K=1}^{R-1}[e_1^K,\dots, e^K_{n_K}]\oplus [e_2^K,\dots, e^K_{n_K}] \oplus\dots \oplus [e^K_{n_K}]$$ 
\be\label{2.55} ~~~~~~~~~ ~~\oplus_{L=0}^{R+1}[\epsilon_1^L,\dots, \epsilon_{m_L}^L]\oplus [\epsilon_2^L,\dots, \epsilon_{m_L}^L] \oplus\dots \oplus [\epsilon_{m_L}^L].\ee
The graph that corresponds to the tridiagonal matrix $K^{IJ}_{\rm{reduced}}=K^{IJ}$ is a connected chain with the decorations shown in the Fig. \ref{f5}.

This reduced plumbing graph $\Gamma_{\rm{reduced}}^p$  represents the same $\mathbb Z$-homology sphere \linebreak $\Sigma (\Gamma_{\rm{reduced}}^p)$ 
as well as the plumbing graph $\Gamma_p$, thus the graph manifold $ M(\Gamma_p)$ is homeomorphic to $\Sigma (\Gamma_{\rm{reduced}}^p)$ and has the opposite orientation; this fact was indicated in \cite{NeumLect}. Hence, we have $M^3_{+}=\Sigma (\Gamma_{\rm{reduced}}^p)=-M(\Gamma_p).$ 

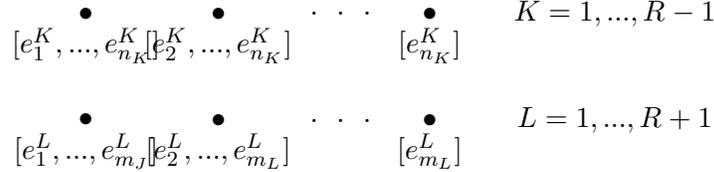
\begin{figure}[h]
\begin{center}

\setlength{\unitlength}{1pt}
\begin{picture}(200,50)

\put(0,40){\makebox(.5,.5){$\bullet$}}
\put(0,28){\makebox(.5,.5){$[e^K_1,...,e^K_{n_K}]$}}

\put(50,40){\makebox(.5,.5){$\bullet$}}
\put(50,28){\makebox(.5,.5){$[e^K_2,...,e^K_{n_K}]$}}

\multiput(85,40)(10,0){3}{.}

\put(130,40){\makebox(.5,.5){$\bullet$}}
\put(130,28){\makebox(.5,.5){$[e^K_{n_K}]$}}

\put(200,40){\makebox(.5,.5){$K=1,...,R-1$}}

\put(0,0){\makebox(.5,.5){$\bullet$}}
\put(0,-12){\makebox(.5,.5){$[e^L_1,...,e^L_{m_J}]$}}

\put(50,-0.5){\makebox(.5,.5){$\bullet$}}
\put(50,-12){\makebox(.5,.5){$[e^L_2,...,e^L_{m_L}]$}}

\multiput(85,-0.5)(10,0){3}{.}

\put(130,-0.5){\makebox(.5,.5){$\bullet$}}
\put(130,-12){\makebox(.5,.5){$[e^L_{m_L}]$}}

\put(200,0){\makebox(.5,.5){$L=1,...,R+1$}}

\end{picture}\\
\vspace*{1.cm}
\caption{A diagonal plumbing graph $\Gamma^p_{\text{diag}}$. (The set of disjoint points) \label{f6}}
\end{center}
\end{figure}

The diagonal matrix $D^{MN}$ corresponds to the plumbing graph $\Gamma^p_{\rm{diag}}$ that is formed by a set of disjoint points decorated as shown in the Fig. \ref{f6}. 
This graph represents a wide set of disjoint (among each other) lens spaces of type $L(p^K_s,q^K_s)$ and $L(a^L_t,b^L_t)$,  
where $p^K_s / q^K_s=-[e_s^K,\dots, e^K_{n_K}]$, $s=1,\dots,n_K$ and 
$a^L_t / b^L_t=-[\epsilon_t^L,\dots, \epsilon_{m_L}^L]$, $t=1,\dots,m_L$.
If, as we propose in Section 3, the matrix $K^{IJ}=K^{IJ}_{\rm{reduced}}$ is interpreted as a coupling constants matrix of a collection of $R$ gauge interactions `switched on' in the large universe (see, for example, the value of cosmological constant in formula (\ref{5.13})), then each element of the diagonal matrix $D^{MN}$ can be associated with the coupling constant of the unique fundamental interaction that exists in corresponding ``baby universe'' splitting from the large universe. In accordance with the direct sum expression for the linking matrix
(\ref{2.54}) and its graph representation in Figs. 5 and 6,
a seven-dimensional universe is described by:
\be\label{2.56}X^4\times\left(M^3_{+} \bigsqcup_{K,s}L(p^K_s,q^K_s)\bigsqcup_{L,t} L(a^L_t,b^L_t)\right),\ee
then the multidimensional space of large universe is $X^4\times M^3_{+}$ (here $X^4$ is the four-dimensional space-time and $\bigsqcup$ denotes the disjoint sum operation). As it will be shown in section 3 the hierarchy of coupling constants of gauge interactions acting in $X^4$  (after Kaluza-Klein reduction procedure) is described by the linking matrix $K^{IJ}$ of the internal space $M^3_{+}$. The diagonal elements of $K^{IJ}$ have the hierarchy of the coupling constants detected in our (large) universe as we show in and Section 5 of this paper (see also \cite{EHMor}). On the other hand, the multidimensional space-times of  ``baby universes'' are formed by the disjoint set of manifolds homeomorphic to $X^{4,K}_s\times L(p^K_s,q^K_s)$  and $X^{4,L}_t\times L(a^L_t,b^L_t)$. Here the four dimensional space-time manifolds $X^{4,K}_s$ and $X^{4,L}_t$ are homeomorphic to $X^4$, but have different scales, since after a dimensional reduction (integrating over lens spaces) there emerge
the corresponding set of coupling constants $[e_s^K,\dots, e^K_{n_K}]$ and $[\epsilon_t^L,\dots, \epsilon_{m_L}^L]$ (only one coupling constant for each disjoint space-time component of type $X^{4,K}_s$ or $X^{4,L}_t$). For this reason we prescribe to the four-dimensional manifolds the additional indexes $~^K_s$ and $~^L_t$. If we associate each of these coupling constants with the ``cosmological'' one defined in the corresponding component, then the scales of them
will be determined by the absolute values of the continued fractions $[e_s^K,\dots, e^K_{n_K}]$ and $[\epsilon_t^L,\dots, \epsilon_{m_L}^L]$. All of these fractions have the absolute values greater than 1, thus the scales of corresponding three-dimensional spatial section will be of Planck or sub-Planck orders. This justify the name ``baby universes'' that we use since these manifolds resemble the objects, which were constructed in \cite{{LavRubTin}, {Colem}, {GiddStrom}}, in order to cancel the cosmological constant by means of topology fluctuation effect. This wormhole approach to the coupling constants problem of nature was
very popular in late 1980s and in early 1990s, see \cite{{Wienb}, {KlebSuss}, {Preskill}, {GonzaDiaz93}}. We are following the spirit, but not the letter of this approach. The fine tuning effect described in section 5 may be induced by the process of ``baby universes'' splitting from the large universe. In Conclusions we shall return to this, ``baby universes'', interpretation, since some results of the sections 3 - 5 are essential in order to make the discussion more profound.

\section{Seven-dimensional BF-model with the Hierarchical Fractional Hall-type Edge States and Topological Mass Generation on the Base of Graph Manifolds } \label{s3}\setcounter{equation}{0}

In this section we construct a rather simple topological gauge model, namely, an Abelian BF-theory on topologically non-trivial seven-dimensional space $X^4\times M_{+}^3$, where internal three-manifold belongs to the class of graph manifolds $M_{+}^3=-M(\Gamma_p)$ (see the notation in formula (\ref{2.38})).
We start with the seven-dimensional BF action
\be\label{3.1} S_7=  m_0 k \int_{X^4\times M_{+}^3}B_3\wedge F_4. \ee   
Here $k$  is a BF coupling constant and $m_0$ is a mass scale parameter.The dimensional reduction down to the four-dimensional space-time $X^4$ is performed by taking the following ansatz for the three- and four-forms $B_3$ and $F_4$:
\be\label{3.2}B_3=\left(B_I\otimes\kappa^I-\frac{1}{m_0 k}F^I_D\otimes\sigma_I\right)+
\left(F_I\otimes\kappa^I+\frac{1}{m_0 k}d H^I_D\otimes\sigma_I\right);  \ee
\be\label{3.3}F_4=\left(F_I\otimes d\kappa^I+\frac{1}{m_0 k}d H^I_D\otimes d\sigma_I\right)=d \left(A_I\otimes d\kappa^I+\frac{1}{m_0 k} H^I_D\otimes d\sigma_I\right).  \ee 
Here $\{\kappa^I|I=1,\dots, R\}$ are $U(1)^R$ connection one forms and $\{\sigma_I|I=1,\dots, R\}$ are their dual one-forms in the sense
\be\label{3.4} \int_{M_{+}^3}\sigma_I\wedge d\kappa^J = \int_{M_{+}^3}d\sigma_I\wedge\kappa^J=\delta_I^J, \ee
see more careful description in Subsection \ref{s2.1}. The two-forms $F_I$, $F^I_D$ and $B_I$ pertain to the de Rham cochain complex $C^2_{dR}(X^4)$ and $H^I_D\in C^1_{dR}(X^4)$ (see more details in \cite{{Zu1}, {Zu2}}). Note that locally $F_I=d A_I$, where $A_I$ are Abelian potentials.

Inserting this ansatz in the action (\ref{3.1}) and performing the integration over the graph manifold $M^3_{+}=-M(\Gamma_p)$ give us a version of $U(1)^R$ BF-model
$$S_4=\int_{X^4}\left( m_0k K^{IJ}B_I\wedge F_J+B_I\wedge d H_D^I-F_I\wedge F^I_D-\frac{1}{m_0k}K_{IJ}F^I_D\wedge d H_D^J+\right.$$
\be\label{3.5}\left. m_0kK^{IJ}F_I\wedge F_J+2F_I\wedge d H_D^I+\frac{1}{m_0k}K_{IJ}d H_D^I\wedge d H_D^J\right),   \ee
where $ K^{IJ}=\int_{M^3_{+}}\kappa^I\wedge d\kappa^{J}$ is the rational linking matrix defined in (\ref{2.38}). The matrix $K_{IJ}$, inverse to the rational linking matrix, is the integer-valued one for the type of graph manifolds considered in this paper (see \cite{SavArt} for more details). The ansatz (\ref{3.2}), (\ref{3.3}) and the action (\ref{3.5}) are invariant under the large collection of gauge transformations:  
\be\label{3.6} A_I\rightarrow A_I+u_I;~~~H_D^I\rightarrow H_D^I-m_0k K^{IJ}u_J;\ee
\be\label{3.7} B_I\rightarrow B_I+v_I;~~~F_D^I\rightarrow F_D^I+m_0k K^{IJ}v_J,\ee
where $u_I$ are arbitrary one-forms and $v_I$ are arbitrary two-forms. These
symmetries will be broken by the choice of the gauge conditions (\ref{3.17}).
The last three terms in (\ref{3.5}) constitute the boundary part (edge states) of the action. The boundary action has the following form
\be\label{3.8}S_{\rm{boundary}}=\int_{\partial X^4}\left( m_0kK^{IJ}A_I\wedge d A_J+2H_D^I\wedge d A_I+\frac{1}{m_0k}K_{IJ}H_D^I\wedge d H_D^J\right),   \ee
that resembles the action of the fractional quantum Hall (FQH) system. If we define the set of background $U(1)$ gauge fields $A^I_{\rm{bg}}=\frac{1}{m_0}H_D^I$, then this action can be read as
\be\label{3.9}S_{\rm{boundary}}=m_0\int_{\partial X^4}\left(kK^{IJ}A_I\wedge d A_J+2A^I_{\rm{bg}}\wedge d A_I+\frac{1}{k}K_{IJ}A^I_{\rm{bg}}\wedge d A^J_{\rm{bg}}\right).\ee
Following to \cite{BerSemSz} we choose $k=\det(K_{IJ})$ and $S_{\rm{boundary}}$ converts (up to common factor $m_0$) in an analogue of FQH system action with $R$ background fields, as it was defined in Section 3 of \cite{BalChanSat}. Using once more \cite{BalChanSat}, we can assume  
\be\label{3.10}A^I_{\rm{bg}}=t^I A_{\rm{ext}}.\ee
Here $t^I$ is a charge vector of the topological Hall fluid \cite{WenZee}. Under these suppositions we get the action for the hierarchical FQH system with unique ``external'' field $A_{\rm{ext}}$:
\be\label{3.11}S_{\rm{FQH}}=m_0\int_{\partial X^4}\left(\tilde{K}^{IJ}A_I\wedge d A_J+2t^I A_{\rm{ext}}\wedge d A_I+\tilde{K}_{IJ}t^I t^J A_{\rm{ext}}\wedge d A_{\rm{ext}}\right),\ee
where $\tilde{K}^{IJ}=\det(K_{IJ})K^{IJ}$ is the tridiagonal integer matrix, which together with the  charge vector $t^I$, characterizes generalized hierarchical Hall states \cite{WenObzor}, and $\tilde{K}_{IJ}=K_{IJ}/\det(K_{IJ})$ is the rational matrix inverse to $\tilde{K}^{IJ}$. The filling factor $\nu=\tilde{K}_{IJ}t^I t^J$ describes the conductivity $\sigma=\nu/2\pi$ of the topological Hall fluid.

With the currents $J_I$ defined as $J_I=(1/2\pi)\ast d A_I$, the equations of motion
\be\label{3.12}\tilde{K}^{IJ}d A_J+t^I d A_{\rm{ext}}=0 \ee
for the action (\ref{3.11}) read
\be\label{3.13}J_I=-(1/2\pi)\tilde{K}_{IJ}t^J\ast d A_{\rm{ext}}.\ee 
Here the Hodge $\ast$ operation is defined with respect to the boundary $\partial X^4$.
Finally we get
\be\label{3.14}t^I\ast J_I=(\nu/2\pi) d A_{\rm{ext}}=\sigma F_{\rm{ext}},\ee
Which is a well known equation of FQH effect. Note that this equation may be obtained also by variation of (\ref{3.11}) with respect to $A_{\rm{ext}}$.
Thus on the non-trivial spatial boundary $\partial X^4$ the action (\ref{3.5}) induces hierarchical FQH effect. 

We can suppose that the bulk part of the action (\ref{3.5})
\be\label{3.15}S_4^{\rm{bulk}}=\int_{X^4}\left( m_0k K^{IJ}B_I\wedge F_J-F_I\wedge F^I_D+B_I\wedge d H_D^I-\frac{1}{m_0k}K_{IJ}F^I_D\wedge d H_D^J\right) \ee
describes a hierarchical system of ``quasiparticles'' in the four-dimensional space-time that is an analogue of fractional topological fluid \cite{{WenKniga}, {ChoMoore}, {MacQiKarZang}}.
We shall show that this action contains the topological mass generation effect, which resembles one investigated in \cite{AllBLah} and permits to describe the topological order phenomenon in dynamical systems possessing ground states with a mass gape for all excitations \cite{Diamant05}. 

Now we will concentrate on the derivation of the classical equations obtained from the action (\ref{3.5}) for the case of compact spatial manifolds without boundaries and we require that fields go to the pure gauge configuration at the infinity in the time direction. Thus the boundary terms do not affect the result. So we can consider only first four terms of this action and rewrite them (up to other boundary term) in the following form   
\be\label{3.16}S_{\rm BF}=\int_{X^4}\left( m_0k K^{IJ}B_I\wedge F_J-F_I\wedge F^I_D-H_I\wedge  H_D^I-\frac{1}{m_0k}K_{IJ}F^I_D\wedge d H_D^J\right), \ee
where $H_I=d B_I$. The rational linking matrix $K^{IJ}$ may be interpreted as a BF {\it coupling constants matrix}, at least the matrix $K^{IJ}$ describes the hierarchy of gauge coupling constants up to dimensionless the scale factor $k$. This interpretation we shall widely used in the following sections. There exists a wide set of Kaluza-Klein models, in which the collections of coupling constants 
are determined mainly by linking (intersection) matrices of internal spaces \cite{{Verl}, {DijVerVonk}, {DijHollSulVa}}.

Let us choose the following gauge conditions :
\be\label{3.17}F^I_D=\frac{1}{2}K^{IJ}\ast F_J;~~~H^I_D=\frac{1}{2}K^{IJ}\ast H_J  \ee
which break the large topological symmetries (\ref{3.6}) and (\ref{3.7}) of the action (\ref{3.16}). (Here the Hodge $\ast$ operation is defined with respect to the four-dimensional space-time $X^4$.) The result reads 
\be\label{3.18}S_{\rm{calibr}}=\int_{X^4}K^{IJ}\left( m_0k B_I\wedge F_J-\frac{1}{2}F_I\wedge \ast F_J-\frac{1}{2}H_I\wedge\ast  H_J-\frac{1}{4m_0k}\ast F_I\wedge d\ast H_J\right). \ee
Note that this action differs from the one which describes the low-energy effective theory of superconductors with topological order given in \cite{Diamant05}, only by the last term. Moreover the action (\ref{3.18}) has two $U(1)^R$ gauge symmetries under the transformations
\be\label{3.19}A_I\rightarrow A_I+\xi_I,~~~~B_I\rightarrow B_I+\eta_I,  \ee
where $\xi_I$ and $\eta_I$ are closed one-forms and two-forms respectively. These transformations give a generalization of gauge symmetries of the topological matter action for topological superconductivity considered in \cite{DiamantI}.

By variation of (\ref{3.18}) with respect to $B_I$ and $A_I$, we obtain 
\be \label{3.20}\left(\square - m^2\right)~F_I=2 m~d\ast H_I;      \ee
\be \label{3.21}\left(\square - m^2\right)~H_I=-2 m~d\ast F_I,      \ee
where $\square=d\ast d\ast +\ast d\ast d$ is the Laplacian and $m=2m_0k$ is a topological mass \cite{Diamant05}. The topological mass plays the role of the gap, characterizing the ground state of topological fluids, and determines the vacuum energy density for low-energy effective field theory, 
which might emerge from the description of small excitations about the established condensate of  topological defects 
\cite{{QuevTrug}, {DiamantI}, {BrazJHEP}}. Following to the Julia-Toulouse approach  \cite{{JulToul}, {Braz0908.0370}}, we need not know the details of the condensate formation process, but it is important that the condensate defines the vacuum and carries the essential information used in the construction of an effective theory. The dynamical equations (\ref{3.20}) and (\ref{3.21}), which we obtained from a formal Kaluza-Klein approach, have the same type as the equation (3.22) in \cite{BrazJHEP}, which follows from the Julia-Toulouse approach in the case when generalized
``electric'' defects condensate.
Note that the gap arises in our model from a topological mechanism (induced by BF term as in \cite{{Diamant05}, {DiamantI}}) and not from local order parameter acquiring a vacuum expectation value. A related mechanism for the topological mass generation is described in \cite{AllBLah} (see also \cite{Braz1202.3798} where the topological BF term arises as a result of condensation of topological defects).

The equations (\ref{3.20}) and (\ref{3.21}) 
are the Klein-Gordon equations for massive fields $F_I$ and $H_I$ with sources. The physical sense of these dynamical equations becomes more clear if we rewrite them in terms of ``electric charge'' currents
\be \label{3.22}J^e_I=\ast H_I \ee
and their dual ``magnetic flux'' currents   
\be \label{3.23}J_m^I=K^{IJ}\ast F_I \ee
analogous of which have been introduced in \cite{{Diamant05}, {DiamantI}}.
The dynamical equations for these generalized currents read
\be \label{3.24}\left(\square - m^2\right)~\ast J_m^I=-2 m~K^{IJ}d J_J^e;      \ee
\be \label{3.25}\left(\square - m^2\right)~\ast J_I^e=-2 m~K_{IJ}d J_m^J.      \ee
These equations tell us that ``magnetic flux'' currents are sources for ``electric charge'' currents and {\it vice versa}. 
Namely these matter currents $J_I^e$ and $J_m^I$ describe the excitations about a topologically ordered (vacuum) state and the effective field theory is formulated for these excitations as a gauge model with the action  
\be\label{3.26}S_{\rm{calibr}}=\frac{1}{2}\int_{X^4}\left(m~K^{IJ} A_I\wedge \ast J^e_J+K_{IJ}J_m^I\wedge \ast J_m^J+K^{IJ}J^e_I\wedge\ast J^e_J-\frac{1}{m} J_m^I\wedge d J_I^e\right), \ee
obtained from (\ref{3.18}). In this action, as well as in (\ref{3.16}), the rational linking matrix $K^{IJ}$ plays the role of hierarchical BF coupling constants matrices.

\section{The Collection of Graph Manifolds as a Base for Description of Coupling Constants Hierarchy and Fine Tuning Effect } \label{s4}
\setcounter{equation}{0}

In this section, we construct a family of the three-dimensional graph manifolds
which play the role of internal spaces whose linking matrices describe
the hierarchical structure of coupling constants and their fine tuning. In particular, a specific matrix reproduces the hierarchy of the experimental low-energy couplings.

\subsection{The principal ensemble of Brieskorn homology spheres: Definitions and preliminary observations}
\label{s4.1}

Since the basic structure blocks of graph manifolds used in this
paper (see Fig. \ref{f3} for corresponding plumbing diagram) are Seifert fibered Brieskorn homology spheres $\Sigma(a_1,a_2,a_3)$, it is appropriate to give here the following definitions \cite{SavB2}. Let $a_1,~a_2,~a_3$ be pairwise relatively prime positive numbers. The Brieskorn homology sphere (Bh-sphere) $\Sigma(\underline{a}):=\Sigma(a_1,a_2,a_3)$ is defined as the link of Brieskorn singularity
\be \label{4.1} \Sigma(\underline{a}):=\Sigma(a_1,
a_2,a_3):=\left\{{z_1}^{a_1}+{z_2}^{a_2} +{z_3}^{a_3}=0\right\}\cap S^5. \ee
Bh-spheres belong to the class
of Seifert fibered homology spheres (Sfh-spheres) \cite{EisNeum}. On each of these
manifolds, there exists a Seifert fibration with unnormalized Seifert invariants $(a_i,b_i)$ subject to $e(\Sigma (\underline{a}))= \sum_{i=1}^{3} b_i/a_i
=-1/a$, where $a=a_1 a_2 a_3$ and $e(\Sigma (\underline{a}))$ is
its rational Euler number, the well known topological invariant of a
Bh-sphere (see Appendix).\\

\noindent{\bf Remark 4.1.} The singular complex algebraic surface ${z_1}^{a_1}+{z_2}^{a_2} +{z_3}^{a_3}=0$
has the canonical orientation which induces the canonical
orientation of the link $\Sigma(\underline{a})$. Note that the Euler number of the Bh-sphere (considered as a simple graph manifold) play the role of $1\times 1$ rational linking matrix, so the canonical orientation of $\Sigma(\underline{a})$ corresponds to the negative definite linking matrix $-1/a$. We shall use always the graph manifolds with positive definite linking matrices, therefore it is necessary to change the orientation of Bh-spheres: $\Sigma_{+}(\underline{a})=-\Sigma(\underline{a})$; thus 
\be \label{4.2}e(\Sigma_{+}(\underline{a}))=- \sum_{i=1}^{3} b_i/a_i=1/a.\ee
   
The Seifert fibration of Bh-sphere is defined by the $S^1$-action
which reads $t(z_1, z_2, z_3)= (t^{\sigma_1}z_1, t^{\sigma_2}z_2,
t^{\sigma_3}z_3)$, where $t\in S^1$, and $\sigma_i= a/a_i$. This
action is fixed-point-free. The only points of $\Sigma(a_1,
a_2,a_3)$ which have a non-trivial isotropy group $\mathbb{Z}_{a_i}$
are those with one coordinate $z_i$ equal to 0 ($i=1,2,3$). The
fiber through such a point is called an exceptional (singular)
fiber of degree $a_i$. All other fibers are called regular
(non-singular). In general, Sfh-spheres $\Sigma(a_1,...,a_n)$ have
$n$ different exceptional fibers with a stabilizer $\mathbb Z_{a_k}$ (the integers $a_k$, $k=1,...,n$ are pairwise relatively prime integers)
and represent special cases of $\mathbb{Z}$-homology spheres \cite{EisNeum} (see also Appendix).

We begin the construction of the principal ensemble of Bh-spheres with definition of a primary sequence. Let
$p_i$ be the $i$th prime number in the set of positive integers
$\mathbb{N}$, {\it e.g.} $p_1=2, ~ p_2=3,\dots, p_9=23,\dots$.
Then the primary sequence of Bh-spheres is defined as
\be \label{4.3} \left\{ \Sigma_{+}(p_{2n},p_{2n+1},q_{2n-1})|n\in
\mathbb{Z}^+\right\}, \ee
where $q_i:=p_1\cdots p_i$.\footnote{If we change the definition of the Bh-spheres primary sequence, then the hierarchy of coupling constants that is
predicted by our model may be other than the experimental one;
see Conclusions for more detailed discussion.} \\

\noindent{\bf Remark 4.2.}
The first terms in this sequence with $n>0$
(which we really use in this section) are $\Sigma(2,3,5)$ (the Poincar\'e homology
sphere), $\Sigma(7,11,30)$, $\Sigma(13,17, 2310)$, and
$\Sigma(19,23,510510)$. We also include in this sequence as its
first term ($n=0$) the usual three-dimensional sphere $S^3$
with Seifert fibration (Sf-sphere) determined by the mapping
$h_{pq}:S^3\rightarrow S^2$, in its turn defined as $h_{pq}(z_1,
z_2)=z^p_1/z^q_2$ \cite{Scott}. Recall that
$S^3=\{(z_1,z_2)||z_1|^2+|z_2|^2=1\}$ and
$z^p_1/z^q_2\in\mathbb{C}\cup\{\infty\}\cong S^2$. In this paper,
we consider the case $p=1$, $q=2$ and denote this Sf-sphere as
$\Sigma(1,2,1)$, {\it i.e.} $p_0=q_{-1}=1$ in (\ref{4.3}). In this
notation, we use two additional units which correspond to two
arbitrary regular fibers. This will enable us to operate with
$\Sigma(1,2,1)$ in the same manner as with other members of the
sequence (\ref{4.3}).\\

Before to proceed with the general consideration, we want to take up a simplified example, which illustrates in what manner the ``cosmological constant'' might appear in our construction. We start with considering the term with $n=4$ in the primary sequence, that is $\Sigma_{+}(19,23,510510)$, that has the Euler number ({\it i.e.} the $1\times 1$ rational linking matrix) 
\be \label{4.4}e\left(\Sigma_{+}\right) =K^{11}= \frac{-5}{19}+\frac{-3}{23}+\frac{200933}{510510}= \frac{1}{223092870}\approx 4.48\times 10^{-9}. \ee    
Now we consider an operation \cite{EM} over the set of Bh-spheres, which is named a derivative of Bh-sphere  and is defined as 
\be \label{4.5}D^{(1)}\Sigma_{+}(\underline{a})=D^{(1)}\Sigma_{+}(a_1,a_2,a_3):=
\Sigma_{+}(a_1,a_2a_3,a+1),\ee
{\it i.e.} the result of this operation is the Bh-sphere with Seifert invariants $a_1^{(1)}=a_1,~a_2^{(1)}=a_2a_3,~a_3^{(1)}=a+1$ and the Euler invariant $e(D^{(1)}\Sigma_{+}(\underline{a}))=1/a^{(1)}$, where $a^{(1)}=
a_1^{(1)}a_2^{(1)}a_3^{(1)}$.
A hint of this operation can be found in Saveliev's paper \cite{SavRuss}.
By induction, we define the $l$-th derivative $D^{(l)}\Sigma_{+}(\underline{a})=
\Sigma_{+}(a_1^{(l)},a_2^{(l)},a_3^{(l)})$ for any $l\in \mathbb N$. Applying 4-th derivative to the Bh-sphere $\Sigma_{+}(19,23,510510)$, we obtain the Bh-sphere 
$D^{(4)}\Sigma_{+}(19,23,510510)$ with Euler number ($1\times 1$ rational linking matrix)  
\be \label{4.6}\begin{array}{l}e^{(4)}=K^{(4)11}= \frac{-5}{19}+
\frac{84986185048470683619597440588949913929045173055232557772918123429}
{322947503184188597754470274238009672930371657609883719537088869030}\\
+\frac{-1}{6136002560499583357334935210522183785677061494587790671204688511571}=\\
-0.\overline{263157894736842105}+\\
0.263157894736842105~263157894736842105~263157894736842105~2631578947370\dots\\
-1. 629725525927064843\times 10^{-67}\approx 2.66\times 10^{-134}.\end{array} \ee 
The consequences of these calculations are following. 

It is possible to interpret the $1\times 1$-matrix $K^{(4)11}$ (formed by the unique Euler number $e^{(4)}$) as a coupling constant of the unique interaction, ``switched on'' in the space-time $X^4$ owing to the presence of three-dimensional internal space, namely, the Bh-sphere $D^{(4)}\Sigma_{+}(19,23,510510)$ (see comments in the end of section \ref{s2.2} as well as BF-actions (\ref{3.16}) and (\ref{3.26})). By its numerical value, $2.66\times 10^{-134}$, this constant may be identify with the cosmological constant in the contemporary universe. In section 5 we shall argue that the constant $K^{11}\approx 4.48\times 10^{-9}$, that is the Euler number of the Bh-sphere $\Sigma_{+}(19,23,510510)$, can be associated with a cosmological constants at the Planck scale of time. Thus the ratio $K^{(4)11}/K^{11}$ might characterize the contemporary cosmological constant in Planck density units. Note that even by these crude estimations, we have  $\Lambda = K^{(4)11}/K^{11} \approx 5.94\times 10^{-126}$, that is less than the empirical bound (\ref{1.1}) only on three orders.

Moreover, the calculations of Euler numbers by the formulas (\ref{4.4}) and (\ref{4.6}) lead to the conclusion that the absolute value of each summand is many orders larger that the resulting Euler number which represents the cosmological constant. This fact simulates the fine tuning effect in the modeling scheme, which is proposed.

Now we shall extend the ensemble of Brieskorn homology spheres in order to include the coupling constants of other gauge interactions, described in terms of BF topological approach. In \cite{EHMor} we showed that the application of derivative to the Bh-spheres belonging to  
the primary sequence (\ref{4.3}) yields a bi-parametric family
of Bh-spheres, whose Euler numbers reproduce fairly well the
experimental hierarchy of {\it dimensionless low-energy coupling} (DLEC)
constants of the fundamental interactions in the real universe.
For the reader's convenience, we concisely reiterate here, in a modified form,
some results from \cite{EHMor}.
 
This bi-parametric family of Bh-spheres is \be \label{4.7}
\left\{D^{(l)}\Sigma_{+}(\underline{a})=\Sigma_{(+)}(a^{(l)}_{1n},a^{(l)}_{2n},a^{(l)}_{3n})= \Sigma_{(+)}(p^{
(l)}_{2n},p^{(l)}_{2n+1},q^{(l)}_{2n-1})|n,l\in
\mathbb{Z}^+\right\}. \ee 
(Note that the collections of Seifert's invariants
$$\left\{a^{(l)}_{1n},a^{(l)}_{2n},a^{(l)}_{3n}\right\}~~~{\rm and}~~~
\left\{p^{ (l)}_{2n},p^{(l)}_{2n+1},q^{(l)}_{2n-1}\right\}$$ are
equivalent up to ordering.) In \cite{{NovaEM}, {EM}} it was shown that to reproduce the hierarchy of the DLEC constants of the known five
fundamental interactions (including the cosmological one), it is
sufficient to restrict values of parameters as $n,l\in\overline{0,4}$ 
and $n-l\geq 0$.\footnote{Hereinafter $\overline{n,m}$ is the integer number interval from $n$ to $m$, where $n,~m \forall \in \mathbb{Z}$.} With this restriction, the Euler numbers of the Bh-spheres family are given in Table~\ref{tab:1}. 

\begin{table*}
\caption{\label{tab:1}Euler numbers of $(n,e)$-family of Sf- and Bh-spheres.}
\begin{center}
\begin{tabular}{|c|l|l|l|l|l|}
\hline
$\hspace*{-.4cm}_{n}$\hspace*{-.2cm}$\diagdown$\hspace*{-.1cm}$^{e}
$\hspace*{-.4cm} & $4$ & $3$ & $2$ & $1$ & 0 \\
\hline0 &  &  &  &  &\hspace*{-.2cm}
$\mathbf{5.0\hspace*{-.1cm}\times 10^{-1}}$\hspace*{-.2cm}
\\\hline 1 &  &  &  &\hspace*{-.2cm}
$3.3\hspace*{-.1cm}\times\hspace*{-.1cm}10^{-2}$ \hspace*{-.2cm}&
\hspace*{-.1cm}$\mathbf{1.1\hspace*{-.1cm}\times\hspace*{-.1cm}10^{-3}}
$\hspace*{-.2cm}\\\hline 2 &  &  &\hspace*{-.2cm}
$4.3\hspace*{-.1cm}\times\hspace*{-.1cm}10^{-4}$\hspace*{-.2cm}
&\hspace*{-.2cm}
$1.9\hspace*{-.1cm}\times\hspace*{-.1cm}10^{-7}$\hspace*{-.2cm}
&\hspace*{-.2cm} $\mathbf{3.5
\hspace*{-.1cm}\times\hspace*{-.1cm}10^{-14}}$\hspace*{-.2cm}
\\\hline 3 &  &
\hspace*{-.1cm}$2.0\hspace*{-.1cm}\times\hspace*{-.1cm}10^{-6}
$\hspace*{-.2cm} &\hspace*{-.2cm}
$3.8\hspace*{-.1cm}\times\hspace*{-.1cm}10^{-12}$
\hspace*{-.2cm}&\hspace*{-.2cm} $1.5
\hspace*{-.1cm}\times\hspace*{-.1cm}10^{-23}$\hspace*{-.2cm}
&\hspace*{-.2cm}
$\mathbf{2.2\hspace*{-.1cm}\times\hspace*{-.1cm}10^{-46}}$\hspace*{-.2cm}
\\\hline 4 &
\hspace*{-.15cm}$4.5\hspace*{-.1cm}\times\hspace*{-.1cm}10^{-9}
$\hspace*{-.2cm} &\hspace*{-.2cm}
$2.0\hspace*{-.1cm}\times\hspace*{-.1cm}10^{-17}$\hspace*{-.2cm}&
\hspace*{-.1cm}$4.0
\hspace*{-.1cm}\times\hspace*{-.1cm}10^{-34}$\hspace*{-.2cm}
&\hspace*{-.2cm} $1.6\hspace*{-.1cm}\times\hspace*{-.1cm}10^{-67}$
\hspace*{-.2cm}&\hspace*{-.2cm}
$\mathbf{2.7\hspace*{-.1cm}\times\hspace*{-.1cm}10^{-134}}$
\hspace*{-.3cm}\\\hline
\end{tabular}
\end{center}
\end{table*}

\begin{table*}
\caption{\label{tab:2} Experimental dimensionless low-energy coupling constants {\it vs.}  Euler numbers {\it vs.} diagonal elements of linking matrix $K^{II}(0)$.}
\begin{center}
\begin{tabular}[c]{|l|l|l|l|l|}\hline
$n$ &$\alpha_{{\rm experimental}}$ &{$e\left(  \Sigma^{(n)}_{\phantom{|}n}\right) $}
 & $K^{II}(0)$ &$I$\\
\hline $0$ & $\alpha_{{\rm st}}\sim 1$   &$0.5$& $9.69\times10^{-1}$ & $1$\\
\hline $1$ &$\alpha_{{\rm em}}=7.30\times10^{-3}$   &$1.07\times10^{-3}$
& $7.21\times10^{-3}$ & $2$\\
\hline $2$ &$\alpha_{{\rm weak}}=3.04\times10^{-12}$ & $3.51\times10^{-14}$
&$1.76\times10^{-12}$&$3$\\
\hline $3$ &$\alpha_{{\rm gr}}=6.86\times10^{-45}$ & $2.17\times10^{-46}$
&$3.68\times10^{-44}$&$4$\\
\hline $4$ &  $\alpha_{{\rm cosm}}=1.48\times10^{-123}$ &$2.70\times10^{-134}$ 
&$2.66\times10^{-134}$&$5$\\
\hline \end{tabular}
\end{center}
~~~ \\
 {\footnotesize
{\bf Notes:} {\bf 1.} The dimensionless constant $\alpha_{\rm st}$ is the strong $\mathop{\rm SU}(3)_c$-gauge coupling constant $\alpha_3$ in the quantum chromodynamics strong coupling regime.
{\bf 2.} The fine structure (electromagnetic) constant is
$\alpha_{\mathrm{em}}=e^2/\hbar c$. {\bf 3.} The dimensionless
weak interaction constant is
$\alpha_{\mathrm{weak}}=(G_{\mathrm{F}}/\hbar c)(m_{\mathrm
e}c/\hbar)^2$, $G_{\mathrm{F}}$ being the Fermi constant
($m_{\mathrm e}$ is electron mass). {\bf 4.} The dimensionless
gravitational coupling constant is
$\alpha_{\mathrm{gr}}=8\pi G m_{\mathrm e}^2/\hbar c$,
$G$ being the Newtonian gravitational constant. {\bf
5.} The dimensionless cosmological constant is $\alpha_{\mathrm{cosm}}=\rho_{\Lambda}=\rho_{\mathrm vacuum}/\rho_{\mathrm Planck}$, {\it i.e.} the energy density of cosmological vacuum in the Planck units.
We select five coupling constants (and their combinations) from 31
dimensionless physical constants required by particle physics
and cosmology that were indicated in \cite{TegAgReWil}, which
characterize the low energy approximation to the extended standard model.}
\end{table*}
To make the comparison with the experimental hierarchy of DLEC
constants easier, we have introduced instead of
$l$, a new parameter $e:=n-l$ which plays the role of ``{\it discrete energy scale parameter}'' (the meaning of this term will be clarified after the formula (\ref{5.7})). Just at $e=0$ ($l=n$) the Euler numbers (see the boldface numbers in Table 1) simulate fairly well the experimental hierarchy of DLEC constants (see two first columns in 
Table 2). This enables us to consider the ensemble of Bh-spheres (\ref{4.7}) at $l=n$ \be \label{4.8} E_{\Sigma}(e=0)=
\left\{\Sigma_{+}(a^{(n)}_{1n},a^{(n)}_{2n},a^{(n)}_{3n})= \Sigma_{+}(p^{(n)}_{2n},p^{(n)}_{2n+1},q^{(n)}_{2n-1}) |n\in\overline{0,4}\right\},
\ee as the collection of three-dimensional internal graph spaces (for the seven-dimensional universes) each of which defines only one interaction in the space-time $X^4$, but with the DLEC constants' hierarchy of the real fundamental interactions. 
Then it is natural to suppose that at $e\in \overline{1,4}$ the ensembles of Bh-spheres
\be \label{4.9} E_{\Sigma}(e)=\left\{\Sigma_{+}(a^{(n-e)}_{1n},a^{(n-e)}_{2n},a^{(n-e)}_{3n})=
\Sigma_{+}(p^{ (n-e)}_{2n},p^{(n-e)}_{2n+1},q^{(n-e)}_{2n-1})|n\in
\overline{e,4}\right\} \ee
consist of basic elements
of internal spaces corresponding to higher densities of vacuum energy. The number of terms in the ensemble $E_{\Sigma}(e)$ coincides with the number of disjoint ``universes'', each of which contains only one fundamental interaction ``switched on'' in the four-dimensional space-time $X^4$ by certain internal space and decrease from five at $e=0$ to one at $e=4$ .\\

\noindent{\bf Remark 4.3.}
It is worth being observed that in
our scheme the five (low energy Abelian) interactions are related to the
first nine prime numbers as 2,3,5,7,11,13,17,19,23. To obtain any new interaction, one has to attach a new pair of prime numbers to the preceding set. For example, taking
the next pair (29,31), we come with the same algorithm to a new
coupling constant of the order of magnitude $\alpha_6\approx
10^{-361}$. Thus our model answers the intriguing question:
How many fundamental interactions may really exist in the
universe? Our model predicts an infinite number of interactions
owing to the infinite succession of prime numbers. We simply cannot
detect too weak interactions beginning with $\alpha_6$ since all
subsequent are even weaker: $\alpha_7\approx 10^{-916}$, {\it
etc.}. This question was discussed with more details in \cite{EHMor}.\\

\subsection{The principal ensemble of Brieskorn homology spheres: the main construction of graph manifolds}
\label{s4.2}
Until now in this section we described the seven-dimensional universes with Bh-spheres as internal spaces. Each of these internal spaces is characterized by $1\times1$ rational linking matrix (Euler number). According to the general interpretation (see Section 3) in the topological BF-model, the linking matrices play the role of the coupling constant matrices, thus any ensemble $E_{\Sigma}(e)$ ($e\in \overline{0,4}$) 
describes the finite number of disjoint ``universes'', each of which contains only one fundamental interaction. 

Our universe is characterized at least by five fundamental low-energy interactions (including cosmological one), therefore in order to model this observational fact, the rational linking matrix $K^{IJ}(e=0)$ of internal space must have the rank $R\geq 5$, that is, we ought to construct a general graph manifold described by the plumbing diagram, shown in the Fig.\ref{f3} with number $R$ of Bh-spheres pasted by plumbing. If we have in mind an unification of interactions, then the linking matrix might depend on the discrete energy parameter $e$ and ${\rm rank} K^{IJ}(e)\geq 1$. Moreover, as a consequence of its interpretation, the rational linking matrix has to be positive definite. This property is also sufficient condition guaranteeing the convergence of the partition function corresponding to the semi-classical quantization of the BF-model, see, for example, \cite{{Verl}, {DijVerVonk}, {EHMor}}.

In order to guarantee the rational linking matrix $K^{IJ}$ to be positive definite, we can ensure the fulfillment of the condition (\ref{2.14}) for the negative definite graph manifold $M(\Gamma_p)$ obtained directly by plumbing of the Bh-spheres (defined as links of singularities in (\ref{4.1})) and then calculate the matrix $K^{IJ}$ for the manifold $M^3_{+}=-M(\Gamma_p)$ by the formula (\ref{2.38}).

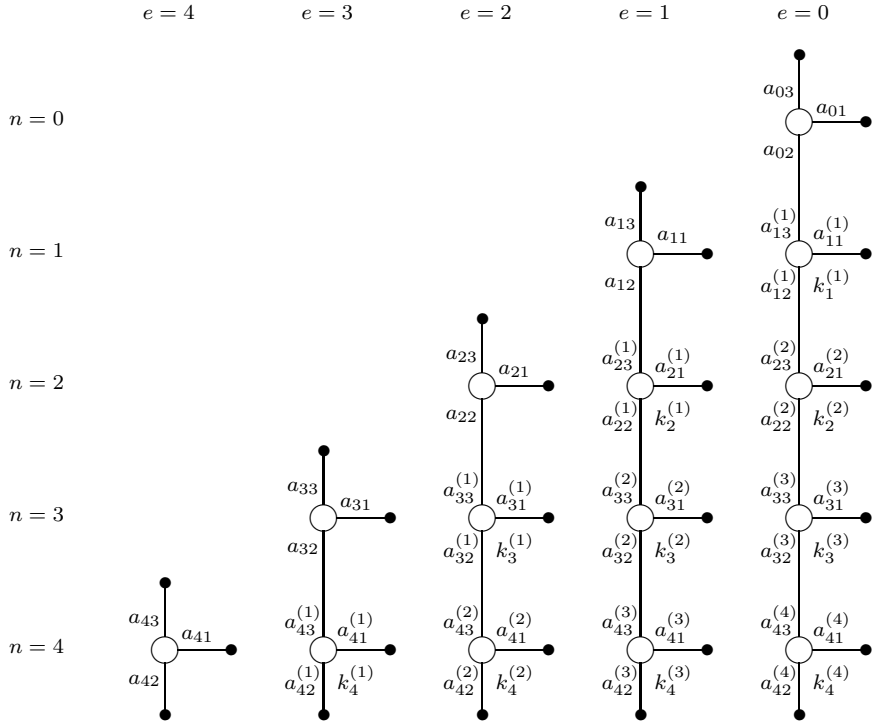
\begin{figure}[h]
\begin{center}
\setlength{\unitlength}{1pt}
\begin{picture}(150,50)

\put(225,-.5){\makebox(.5,.5){$\bullet$}}
\put(225,-50.5){\makebox(.5,.5){$\bullet$}}
\put(225,-100.5){\makebox(.5,.5){$\bullet$}}
\put(225,-150.5){\makebox(.5,.5){$\bullet$}}
\put(225,-200.5){\makebox(.5,.5){$\bullet$}}
\put(200,25){\makebox(.5,.5){$\bullet$}}
\put(200,-225){\makebox(.5,.5){$\bullet$}}

\put(211,3){\makebox(3,3){\scriptsize $a_{01}$}}
\put(191,10){\makebox(3,3){\scriptsize $a_{03}$}}
\put(191,-13){\makebox(3,3){\scriptsize $a_{02}$}}

\put(211,-43){\makebox(3,3){\scriptsize $a^{(1)}_{11}$}}
\put(191,-40){\makebox(3,3){\scriptsize $a^{(1)}_{13}$}}
\put(191,-63){\makebox(3,3){\scriptsize $a^{(1)}_{12}$}}

\put(211,-93){\makebox(3,3){\scriptsize $a^{(2)}_{21}$}}
\put(191,-90){\makebox(3,3){\scriptsize $a^{(2)}_{23}$}}
\put(191,-113){\makebox(3,3){\scriptsize $a^{(2)}_{22}$}}

\put(211,-143){\makebox(3,3){\scriptsize $a^{(3)}_{31}$}}
\put(191,-140){\makebox(3,3){\scriptsize $a^{(3)}_{33}$}}
\put(191,-163){\makebox(3,3){\scriptsize $a^{(3)}_{32}$}}

\put(211,-193){\makebox(3,3){\scriptsize $a^{(4)}_{41}$}}
\put(191,-190){\makebox(3,3){\scriptsize $a^{(4)}_{43}$}}
\put(191,-213){\makebox(3,3){\scriptsize $a^{(4)}_{42}$}}

\put(211,-63){\makebox(3,3){\scriptsize $k^{(1)}_{1}$}}
\put(211,-113){\makebox(3,3){\scriptsize $k^{(2)}_{2}$}}
\put(211,-163){\makebox(3,3){\scriptsize $k^{(3)}_{3}$}}
\put(211,-213){\makebox(3,3){\scriptsize $k^{(4)}_{4}$}}

\put(205,0){\line(1,0){20}}
\put(200,5){\line(0,1){20}}
\put(200,-5){\line(0,-1){20}}
\put(200,0){\circle{10}}
\put(199.5,-.50){\makebox(.5,.5)}

\put(205,-50){\line(1,0){20}}
\put(200,-45){\line(0,1){20}}
\put(200,-55){\line(0,-1){20}}
\put(200,-50){\circle{10}}
\put(199.5,-50.50){\makebox(.5,.5)}

\put(205,-100){\line(1,0){20}}
\put(200,-95){\line(0,1){20}}
\put(200,-105){\line(0,-1){20}}
\put(200,-100){\circle{10}}
\put(199.5,-100.50){\makebox(.5,.5)}

\put(205,-150){\line(1,0){20}}
\put(200,-145){\line(0,1){20}}
\put(200,-155){\line(0,-1){20}}
\put(200,-150){\circle{10}}
\put(199.5,-150.50){\makebox(.5,.5)}

\put(205,-200){\line(1,0){20}}
\put(200,-195){\line(0,1){20}}
\put(200,-205){\line(0,-1){20}}
\put(200,-200){\circle{10}}
\put(199.5,-200.50){\makebox(.5,.5)}

\put(165,-50.5){\makebox(.5,.5){$\bullet$}}
\put(165,-100.5){\makebox(.5,.5){$\bullet$}}
\put(165,-150.5){\makebox(.5,.5){$\bullet$}}
\put(165,-200.5){\makebox(.5,.5){$\bullet$}}
\put(140,-25){\makebox(.5,.5){$\bullet$}}
\put(140,-225){\makebox(.5,.5){$\bullet$}}

\put(151,-45){\makebox(3,3){\scriptsize $a_{11}$}}
\put(131,-40){\makebox(3,3){\scriptsize $a_{13}$}}
\put(131,-63){\makebox(3,3){\scriptsize $a_{12}$}}

\put(151,-93){\makebox(3,3){\scriptsize $a^{(1)}_{21}$}}
\put(131,-90){\makebox(3,3){\scriptsize $a^{(1)}_{23}$}}
\put(131,-113){\makebox(3,3){\scriptsize $a^{(1)}_{22}$}}

\put(151,-143){\makebox(3,3){\scriptsize $a^{(2)}_{31}$}}
\put(131,-140){\makebox(3,3){\scriptsize $a^{(2)}_{33}$}}
\put(131,-163){\makebox(3,3){\scriptsize $a^{(2)}_{32}$}}

\put(151,-193){\makebox(3,3){\scriptsize $a^{(3)}_{41}$}}
\put(131,-190){\makebox(3,3){\scriptsize $a^{(3)}_{43}$}}
\put(131,-213){\makebox(3,3){\scriptsize $a^{(3)}_{42}$}}

\put(151,-113){\makebox(3,3){\scriptsize $k^{(1)}_{2}$}}
\put(151,-163){\makebox(3,3){\scriptsize $k^{(2)}_{3}$}}
\put(151,-213){\makebox(3,3){\scriptsize $k^{(3)}_{4}$}}

\put(145,-50){\line(1,0){20}}
\put(140,-45){\line(0,1){20}}
\put(140,-55){\line(0,-1){20}}
\put(140,-50){\circle{10}}
\put(139.5,-50.50){\makebox(.5,.5)}

\put(145,-100){\line(1,0){20}}
\put(140,-95){\line(0,1){20}}
\put(140,-105){\line(0,-1){20}}
\put(140,-100){\circle{10}}
\put(139.5,-100.50){\makebox(.5,.5)}

\put(145,-150){\line(1,0){20}}
\put(140,-145){\line(0,1){20}}
\put(140,-155){\line(0,-1){20}}
\put(140,-150){\circle{10}}
\put(139.5,-150.50){\makebox(.5,.5)}

\put(145,-200){\line(1,0){20}}
\put(140,-195){\line(0,1){20}}
\put(140,-205){\line(0,-1){20}}
\put(140,-200){\circle{10}}
\put(139.5,-200.50){\makebox(.5,.5)}

\put(105,-100.5){\makebox(.5,.5){$\bullet$}}
\put(105,-150.5){\makebox(.5,.5){$\bullet$}}
\put(105,-200.5){\makebox(.5,.5){$\bullet$}}
\put(80,-75){\makebox(.5,.5){$\bullet$}}
\put(80,-225){\makebox(.5,.5){$\bullet$}}

\put(91,-96){\makebox(3,3){\scriptsize $a_{21}$}}
\put(71,-90){\makebox(3,3){\scriptsize $a_{23}$}}
\put(71,-113){\makebox(3,3){\scriptsize $a_{22}$}}

\put(91,-143){\makebox(3,3){\scriptsize $a^{(1)}_{31}$}}
\put(71,-140){\makebox(3,3){\scriptsize $a^{(1)}_{33}$}}
\put(71,-163){\makebox(3,3){\scriptsize $a^{(1)}_{32}$}}

\put(91,-193){\makebox(3,3){\scriptsize $a^{(2)}_{41}$}}
\put(71,-190){\makebox(3,3){\scriptsize $a^{(2)}_{43}$}}
\put(71,-213){\makebox(3,3){\scriptsize $a^{(2)}_{42}$}}

\put(91,-163){\makebox(3,3){\scriptsize $k^{(1)}_{3}$}}
\put(91,-213){\makebox(3,3){\scriptsize $k^{(2)}_{4}$}}

\put(85,-100){\line(1,0){20}}
\put(80,-95){\line(0,1){20}}
\put(80,-105){\line(0,-1){20}}
\put(80,-100){\circle{10}}
\put(79.5,-100.50){\makebox(.5,.5)}

\put(85,-150){\line(1,0){20}}
\put(80,-145){\line(0,1){20}}
\put(80,-155){\line(0,-1){20}}
\put(80,-150){\circle{10}}
\put(79.5,-150.50){\makebox(.5,.5)}

\put(85,-200){\line(1,0){20}}
\put(80,-195){\line(0,1){20}}
\put(80,-205){\line(0,-1){20}}
\put(80,-200){\circle{10}}
\put(79.5,-200.50){\makebox(.5,.5)}

\put(45,-150.5){\makebox(.5,.5){$\bullet$}}
\put(45,-200.5){\makebox(.5,.5){$\bullet$}}
\put(20,-125){\makebox(.5,.5){$\bullet$}}
\put(20,-225){\makebox(.5,.5){$\bullet$}}

\put(31,-146){\makebox(3,3){\scriptsize $a_{31}$}}
\put(11,-140){\makebox(3,3){\scriptsize $a_{33}$}}
\put(11,-163){\makebox(3,3){\scriptsize $a_{32}$}}

\put(31,-193){\makebox(3,3){\scriptsize $a^{(1)}_{41}$}}
\put(11,-190){\makebox(3,3){\scriptsize $a^{(1)}_{43}$}}
\put(11,-213){\makebox(3,3){\scriptsize $a^{(1)}_{42}$}}

\put(31,-213){\makebox(3,3){\scriptsize $k^{(1)}_{4}$}}

\put(25,-150){\line(1,0){20}}
\put(20,-145){\line(0,1){20}}
\put(20,-155){\line(0,-1){20}}
\put(20,-150){\circle{10}}
\put(19.5,-150.50){\makebox(.5,.5)}

\put(25,-200){\line(1,0){20}}
\put(20,-195){\line(0,1){20}}
\put(20,-205){\line(0,-1){20}}
\put(20,-200){\circle{10}}
\put(19.5,-200.50){\makebox(.5,.5)}

\put(-15,-200.5){\makebox(.5,.5){$\bullet$}}
\put(-40,-175){\makebox(.5,.5){$\bullet$}}
\put(-40,-225){\makebox(.5,.5){$\bullet$}}

\put(-29,-196){\makebox(3,3){\scriptsize $a_{41}$}}
\put(-49,-190){\makebox(3,3){\scriptsize $a_{43}$}}
\put(-49,-213){\makebox(3,3){\scriptsize $a_{42}$}}

\put(-35,-200){\line(1,0){20}}
\put(-40,-195){\line(0,1){20}}
\put(-40,-205){\line(0,-1){20}}
\put(-40,-200){\circle{10}}
\put(-40.5,-200.50){\makebox(.5,.5)}

\put(200,40){\makebox(3,3){\scriptsize $e=0$}}
\put(140,40){\makebox(3,3){\scriptsize $e=1$}}
\put(80,40){\makebox(3,3){\scriptsize $e=2$}}
\put(20,40){\makebox(3,3){\scriptsize $e=3$}}
\put(-40,40){\makebox(3,3){\scriptsize  $e=4$}}

\put(-90,0){\makebox(3,3){\scriptsize $n=0$}}
\put(-90,-50){\makebox(3,3){\scriptsize $n=1$}}
\put(-90,-100){\makebox(3,3){\scriptsize $n=2$}}
\put(-90,-150){\makebox(3,3){\scriptsize $n=3$}}
\put(-90,-200){\makebox(3,3){\scriptsize $n=4$}}

\end{picture}\\
\vspace*{8.5cm} 
\caption{The plumbing diagrams of different states of the internal space. \label{f7}}
\end{center}
\end{figure}

The natural plumbing diagrams corresponding to internal spaces for the values $e\in\overline{0,4}$ of discrete energy scale parameter are such as  presented in Fig. \ref{f7}. This triangular diagram corresponds to the triangular Table 1 in the sense that to each Bh-sphere (plumbed according to the vertical edges in Fig.\ref{f7}) the appropriate Euler number is associated in the Table 1. The number of Bh-spheres corresponding to the discrete energy parameter $e$ coincides with the rank $R(e)$ of the rational linking matrix $K^{IJ}(e)$ calculated for the vertical connected subdiagrams $\Delta_p(e)$, and consequently coincides with the number of gauge interactions defined by the internal space formed by the graph manifold $M^3_{+}(e)=-M(\Gamma_p(e))$. The number of these interactions decrease from five at $e=0$ to one at $e=4$. It is possible to interpret this fact  as an unification of  fundamental interactions in our model.

In order to construct the internal spaces $M^3(e)$ with properties indicated in the second and third paragraphs of this subsection the modification of the notion of derivative for Bh-sphere is necessary to introduce, namely, we shall use a new concept of $k$-derivative. The $k$-derivative of each Bh-spheres
$\Sigma(a_1,a_2,a_3)$ is another
Bh-sphere \be \label{4.10} \Sigma_{k^{(1)}}(a^{(1)}_1,a^{(1)}_2,
a^{(1)}_3)=\Sigma(a_1,a_2a_3,k^{(1)}a+1), \ee {\it i.e.} it is the
Bh-sphere with Seifert invariants \be \label{4.11} a^{(1)}_1=a_1,
a^{(1)}_2=a_2a_3, a^{(1)}_3=k^{(1)}a+1, \ee where $a=a_1a_2a_3$, $k^{(1)}
\in\mathbb{N}$. The upper index in the parentheses means a single
application of $k$-derivative. A repeated application of
this operation yields still another Bh-sphere \be \label{4.12}
\Sigma_{k^{(1)}k^{(2)}}(a^{(2)}_1,a^{(2)}_2, a^{(2)}_3)=
\Sigma(a_1,a_2a_3(k^{(1)}a+1),k^{(2)}a(k^{(1)}a+1)+1), \ee where $k^{(2)}\in
\mathbb{N}$; in general, $k^{(2)}\neq k^{(1)}$. The $l$-fold application
of the $k$-derivative gives again an Bh-sphere, $\Sigma_{
k^{(1)}\dots k^{(l)}}(a^{(l)}_1,a^{(l)}_2, a^{(l)}_3)$ whose invariants
are found by induction from the invariants $a^{(l-1)}_1,a^{
(l-1)}_2,a^{(l-1)}_3$ with arbitrary $k^{(l)}\in\mathbb{N}$. Note
that the least Seifert invariant does not change under
$k$-derivatives ($a^{(l)}_1=a_1$ for any $l=1,2,\dots$), while
the two other Seifert invariants depend on the order
(multiplicity) of the $k$-derivative fulfillment. 

Now we are ready to construct the internal spaces $M_{+}^3(e)$ for all plumbing subdiagrams
shown in the Fig. \ref{f7}. First of all we note that each Bh-sphere in this figure is characterized by three parameters $n,~l$ and $e$. Only two of them are independent, since $e=n-l$. The Seifert invariants of the plumbed Bh-spheres are presented in the form $a^{(l)}_{ni}$, but it is possible to represent them in other manner $a_i^I(e)$ , with the condition $I=l+1$, that is, the same notation as in Fig. 3 with the additional parameter $e$, numerating the disjoint plumbing diagrams as in Fig. 7. The parameter $k$ of $k$-derivate is provided with two indexes, namely, $k^{(l)}_n$. 
Then, the total diagram in Fig. \ref{f7} consists of five
connected plumbing subdiagrams, each of which has the form of the plumbing diagram shown in the Fig. \ref{f3}. Naturally, there exists an ambiguity
in the plumbing operation (related to this type of diagrams). This
diagram contains fifteen nodes, each of them having
three adjacent edges. Thus one can glue $3^{15}$ different graph manifolds of type
$$\bigsqcup_{e=0}^4 M_{+}^3(e),$$ with five connected graph submanifolds.
Moreover, there is an infinite set of integers $k^{(l)}_n$
which guarantee positive definiteness of the linking matrices
of respective graph manifolds. It is however possible to fix a unique
gluing procedure imposing a minimality condition on the
coefficients $k^{(l)}_n$ at each level of realization of the
$k$-derivative. In particular, this condition immediately yields
a conclusion that the vertices corresponding to minimal
Seifert invariants ($a^{(l)}_{n1}$) remain free (not subjected to
plumbing). Applying the $k$-derivative to all Bh-spheres of the
primary subsequence
\be \label{4.13} \left\{\Sigma_{+}(p_{2n},p_{2n+1},q_{2n-1})=\Sigma_{+}(a_{1n},a_{2n},a_{3n}) |n\in\overline{1,4}\right\}, \ee
of the primary sequence (\ref{4.3}), we find the minimal
$k^{(1)}_n$ for which the conditions (\ref{2.14}) are satisfied.
In the case under consideration these conditions read  
$$a_{n2}a^{(1)}_{n+1,3}>a_{n1}a_{n3}a^{(1)}_{n+1,1}a^{(1)}_{n+1,2},$$ for 
$n\in\overline{0,3}$. We omit the superindex $(l)$, when $l=0$. Thus we unambiguously fixed the collection of the first-level Bh-spheres, {\it i.e.} those
with the parameter $l=1$: \be \label{4.14}
\left\{\Sigma_{k^{(1)}_n}
(a^{(1)}_{1n},a^{(1)}_{2n},a^{(1)}_{3n})|n\in\overline{1,4}
\right\}. \ee Now we execute the first plumbing procedure (along
the upper vertical edges between nodes in Fig. \ref{f7}): \be
\label{4.15} \Sigma_{+}(a_{n1},a_{n2}, a_{n3})\frac{~}{a_{n2}~~~~~~a^1_{n+1,3}}
\Sigma_{k^{(1)}_{n+1}} (a^{(1)}_{n+1,1},a^{(1)}_{n+1,2},
a^{(1)}_{n+1,3}) \ee where $n\in\overline{0,3},$ 
that is, the plumbing is performed along the exceptional fibers with the Seifert invariants $a_{n2}$ and $a^{(1)}_{n+1,3}$ (see Appendix).
The same algorithm is applied to determine the
collection of the second-level Bh-spheres, and further by
induction the $l$th-level Bh-spheres: \be \label{4.16}
\left\{\Sigma_{k^{(1)}_n\dots k^{(l)}_n}
(a^{(l)}_{n1},a^{(l)}_{n2},a^{(l)}_{n3})|k^{(1)}_n=\cdots =k^{(l-1)}_n=1, n\in\overline{l,4}\right\}, \ee
and in each step, the plumbing operation is executed 
according to the diagram in Fig. \ref{f7}. Note that the condition $k^{(1)}_n=\cdots =k^{(l-1)}_n=1$ has been imposed since 1 is the absolute minimum of coefficients $k^{(s)}_n$ (see the formulas (\ref{4.10}) and (\ref{4.11})). The coefficient $k^{(l)}_n$ has to be chosen  as the minimal number which satisfy the inequality of type 
\be \label{4.17}a^{(l)}_{n2}a^{(l+1)}_{n+1,3}>
a^{(l)}_{n1}a^{(l)}_{n3}a^{(l+1)}_{n+1,1}a^{(l+1)}_{n+1,2};\ee
to do this it is sufficient to take only $k^{(l)}_n\geq 1$.

Consequently, we obtain the five connected subdiagrams
$\Delta_p^{\rm min}(e)$ where superscript $~^{\rm min}$ corresponds to the use of
$k$-derivative with {\it minimization} of parameters $k^{(l)}_n$ at
each step. According to the procedure described in the section
\ref{s2}, the corresponding plumbing graphs
$\Gamma^{\rm min}_p(e)$ can be constructed. These graphs codify the positive
definite graph manifolds $M_{+}^{\rm min}(e)$, $e\in\overline{0,4}$, which
are interpreted as the internal manifolds corresponding to
different values of discrete energy scale parameter $e$. From now we will omit (for brevity) the superscript $~^{\rm min}$, since we don't use other type of internal spaces.

\section{Modeling of Coupling Constants Hierarchy and Fine Tuning Effect by Means of Linking Matrix} \label{s5}
\setcounter{equation}{0}
In this section we try to model the coupling constants hierarchy and the fine tuning effect in terms of linking matrices properties. For this purpose we give the results of exact numerical calculations of the linking matrices $K^{IJ}(e)$ for the components $M_{+}(e)$, $e\in\overline{0,4}$ of the graph manifold, which corresponds to the plumbing diagram shown in Fig. \ref{f7}.
For $e=4$ the graph manifold $M_{+}(4)$ consists of only one Bh-sphere   
$\Sigma_{+}( 19,23,510510)$, and its its $1\times1$ rational linking matrix coincide with the Euler number 
\be \label{5.1}\begin{array}{l}K^{11}(4)= \frac{-5}{19}+\frac{200933}{510510}+\frac{-3}{23}= \frac{1}{223092870}=\\-0.26315789473684\dots+0.39359268182798\dots-0.1304347826087\dots
\approx4.48\times 10^{-9}.\end{array}\ee 
already presented with lesser detailing, as an example, in (\ref{4.4}). 

For $e=3$ the graph manifold $M_{+}(3)$ is the result of plumbing of two Bh-spheres
$\Sigma_{+}( 13,2310,17) \frac{~~~~~}{~~~~~} \Sigma_{+}( 19,11741730,223092871) $
along the exceptional fibers with the Seifert invariants $a_{32}=2310$ and $a_{43}^{(1)}=223092871$. The rational linking matrix has the following nonzero elements: \be\label{5.2}\begin{array}{l}
K^{11}(3)=\frac{4}{13}+\frac{-2}{17}+\frac{-88567869829}{466041007740}=\\ 
0.3076923077\dots -0.1176470588\dots -0.1900430828\dots\approx
2.17\times 10^{-6};\\
K^{22}(3)=\frac{-5}{19}+\frac{3089929}{11741730}+\frac{-2089}{466041007740}=
\frac{1}{45008842404505980}=\\
-0.2631578947\dots +0.2631578992\dots-4.482438166\times 10^{-9}\approx
2.22\times 10^{-17};\\ 
K^{12}(3)=-\frac{1}{466041007740}\approx-2.15\times 10^{-12}.\end{array}\ee 
For the element $K^{11}(3)$, as well as for $K^{22}(3)$, the modeling of the fine tuning effect is very visible.

For $e=2$ the graph manifold $M_{+}(3)$ is the result of plumbing of three Bh-spheres
$$\begin{array}{l}\Sigma_{+}( 7,30,11)\frac{~~~}{~~~}\Sigma_{+}(13,39270,1531531)\frac{~~~}{~~~}\\
\Sigma_{+}( 19,2619496256206830,25283377864908323161).\end{array}$$ 
The rational linking matrix has the following nonzero elements: 
\be \label{5.3}\begin{array}{l}
K^{11}(2)=
\frac{1}{7}+\frac{1}{11}+\frac{-1531537}{6636660}=\\
0.1428571429\dots
+0.\overline{09}-0.2307692424\dots
\approx2.99\times 10^{-3};\\
K^{22}(2)=
\frac{4}{13}+\frac{-13}{6636660}+ \frac{-601177010295723703913}{1953837726068319020160}=\\
0.3076923077\dots -1.958816634\times 10^{-6}
-0.3076903482\dots\approx6.58\times 10^{-10};\\
K^{33}(2)=
\frac{-5}{19}+ \frac{689341120054429}{2619496256206830}+\frac{-39257}{1953837726068319020160}=\\
-0.2631578947\dots +0.2631578947\dots-2.009225202\times 10^{-17} \approx
4.04\times 10^{-34};\\
K^{12}(2)=-\frac{1}{6636660}\approx-1.51\times 10^{-7};\\
K^{23}(2)=-\frac{1}{1953837726068319020160}\approx-5.12\times 10^{-22}.\end{array}\ee
For the diagonal elements $K^{22}(2)$ and $K^{33}(2)$, the modeling of the fine tuning effect is very clear.

For $e=1$ the graph manifold $M_{+}(1)$ is the result of plumbing of four Bh-spheres
$$\begin{array}{l}\Sigma_{+}(2,5,3)\frac{~~~}{~~~}\Sigma_{+}(7,330,4621) \frac{~~~}{~~~}\Sigma_{+}(13,20047766970,\allowbreak 25801476090391)\frac{~~~}{~~~}\\\Sigma_{+} (19,130373452089350371408191290535930,\\
~~~~~~41446763406821197873635276791696434441). \end{array}$$ 
The rational linking matrix has the following nonzero elements:
\be \label{5.4}\begin{array}{l}
K^{11}(1)=
\frac{-1}{2}+\frac{1}{3}+\frac{2311}{9245}= -0.5+0.333\,333\,333\,3\
+0.249\,972\,958\,4 \approx8.33\times 10^{-2};\\
K^{22}(1)=
\frac{1}{7}+\frac{-4}{9245}+\frac{-11988564648107}{84180573507360}=\\ 0.1428571429-4.326663061\times 10^{-4} -0.1424148607 \approx9.62\times 10^{-6};\\
K^{33}(1)=
\frac{4}{13}+\frac{-323}{84180573507360}+\frac{-
15280072342108259958559713567107013122386613}{
49660235112471118762743483357286701277802160}=\\
0.3076923077-3.836989777\times 10^{-12}-0.3076923077\approx
2.50\times 10^{-21};\\
K^{44}(1)=
\frac{-5}{19}+\frac{34308803181407992475839813298929}{130373452089350371408191290535930}+\\
\frac{-20047766957}{49660235112471118762743483357286701277802160}=\\
-0.2631578947\dots+0.2631578947\dots-4.036985913\times 10^{-34} \approx1.63\times 10^{-67};\\
K^{12}(1)=-\frac{1}{9245}\approx-1.08\times 10^{-4};\\
K^{23}(1)=-\frac{1}{84180573507360}\approx-1.19\times 10^{-14};\\
K^{34}(1)=-\frac{1}{49660235112471118762743483357286701277802160}
\approx-2.01\times 10^{-44}.\end{array}\ee
An analysis of fine tuning for $K^{44}(1)$ will be given in (\ref{5.14}).

For $e=0$ the graph manifold $M_{+}(0)$ is the result of plumbing of five Bh-spheres
\be\label{5.5}\begin{array}{l}
M_{+}(0)=\Sigma_{+}(1,2,1)\frac{~~~}{~~~}\Sigma_{+}(2,15,31)\frac{~~~}{~~~}\Sigma_{+}(7,69330,2426551) \frac{~~~}{~~~}\\
\Sigma_{+}( 13,5224868486304546518670,16709129419201939766706661) \frac{~~~}{~~~}\\
\Sigma_{+}(19,322947503184188597754470274238009672930371657609883719537088869030,\\ 255104306452770178081199931377459790889523831637487397155334924868522751).\end{array}
\ee
The rational linking matrix has the following nonzero elements:
\be \label{5.6}\begin{array}{l}K^{11}(0)=
\frac{0}{1}+\frac{0}{1}+\frac{31}{32}
\approx0+0+0.96875\approx0.96875\\
K^{22}(0)=
\frac{1}{2}+\frac{-1}{32}+\frac{-32030467}{69399345}=0.5-0.03125-0.4615384626\approx
7.21\times 10^{-3}\\

K^{33}(0)=
\frac{1}{7}+\frac{-13}{69399345}+
\frac{-7399970787416156625299392607}{51799863435203418336200490960}\\
=0.1428571429-1.873216527\times 10^{-7} -0.1428569555\approx1.76\times 10^{-12}\\

K^{44}(0)=
\frac{4}{13}+\frac{-762623}{51799863435203418336200490960}+\\ \frac{-
9864555818534855297888748326721335173456647233646233412379742243185343546551763058385313}{
32059806410238279718139966062484464209573437243152889136180581555726013473961031111880160}\\
=0.3076923077\dots-1.472249055\dots \times 10^{-23}
-0.3076923077\dots\approx3.68\times 10^{-44}\\

K^{55}(0)=\frac{-5}{19}+\frac{84986185048470683619597440588949913929045173055232557772918123429}
{322947503184188597754470274238009672930371657609883719537088869030}+\\
\frac{-5224868486304546518657}{
32059806410238279718139966062484464209573437243152889136180581555726013473961031111880160}
\\ =-0.2631578947\dots +0.2631578947\dots
-1.629725526\dots\times 10^{-67}
\approx2.66\times 10^{-134}\\
K^{12}(0)=-\frac{1}{32}\approx-0.03125\\
K^{23}(0)=-\frac{1}{69399345}\approx-1.44\times 10^{-8}\\
K^{34}(0)=-\frac{1}{51799863435203418336200490960}\approx
-1.93\times 10^{-29}\\
K^{45}(0)=\\-\frac{1}{
32059806410238279718139966062484464209573437243152889136180581555726013473961031111880160}\\
\approx -3.12\times 10^{-89}. \end{array}\ee
More detail discussion of fine tuning for the ``low-energy cosmological constant'', $K^{55}(0)$, will be given in (\ref{5.12}).  
 
We begin the physical interpretation from the discussion of the graph manifold $M_{+}(0)$ and its rational linking matrix $K^{IJ}(0)$ which, according with calculations presented in (\ref{5.6}), is convenient to represent as
\be\label{5.7}K^{IJ}(0)=\left(
\begin{array}{lllll}
{\bf 9.69\times 10^{\text{-1}}} & -3.13\times 10^{\text{-2}}
& 0 & 0 & 0 \\
-3.13\times 10^{\text{-2}} & {\bf 7.21\times 10^{\text{-3}}}
& -1.44\times 10^{\text{-8}} & 0 & 0 \\
0 &- 1.44\times 10^{\text{-8}} & {\bf 1.76\times 10^{\text{-12}}}
& -1.93\times 10^{\text{-29}} & 0\\
0 & 0 & -1.93\times 10^{\text{-29}} & {\bf 3.68\times 10^{\text{-44}}}
& -3.12\times 10^{\text{-89}} \\
0 & 0 & 0 & -3.12\times 10^{\text{-89}} & {\bf 2.66\times
10^{\text{-134}}}.
\end{array}
\right)\ee
Recall that in the BF-model constructed in the section 3, a rational linking matrix $K^{IJ}$ of a graph manifold $M^3_{+}$  is interpreted as matrix that describes the hierarchy of gauge coupling constants, which appears as a result of the reduction from the seven-dimensional manifold $X^4\times M^3_{+}$ down to the four-dimensional space-time $X^4$.
The diagonal elements of the matrix $K^{IJ}(0)$ have a hierarchy very closed to the one of the dimensionless low energy coupling (DLEC) constants, as it is shown in the last column of table 2. It is natural to suppose that diagonal elements of the other
rational linking matrices $K^{IJ}(e)$, $e\in
\overline{1,4}$ (as well as their eigenvalues, see  \cite{EHMor} ) 
simulate hierarchy of the vacuum-level coupling
constants of the fundamental interactions
acting in the states characterized by
higher densities of vacuum energy. (According to this hypothesis we refer to $e$ as a {\it discrete energy scale parameter}.)
The sequence of these states correspond to the successive changes of the topological structure of extra dimensional space $M^3_{+}(e)$. These internal space topology transformations induce the changes of number of Abelian gauge fields $A_I(e)$ (where $I=1,\dots,R(e)={\rm rank} K^{IJ}(e)$) and of their coupling constants hierarchies described by matrices $K^{IJ}(e)$.
Thus our model includes a certain ``unification scheme''  
of Abelian  interactions of BF type (described the set of hierarchical topological  fluids mentioned in subsection 2.2) in the following sense.
The linking matrix $K^{IJ}(1)$ has
the rank 4 and hence it describes the state with four
interactions. This state can be associated with
the density of vacuum energy
which corresponds to the topological
structure of the internal space described by the graph manifold $M^3_{+}(1)$. 
But it would be too speculative to connect (at least directly)  this ``unification'' with the unification within the framework of standard model extended by gravitation, see \cite{TegAgReWil}, 
since in our model five low-energy interactions (having coupling constants hierarchy characterized by matrix  $K^{IJ}(0)$)  are replaced by four interactions with rather different coupling constants hierarchy described by matrix $K^{IJ}(1)$:
\be\label{5.8}K^{IJ}(1)=\left(
\begin{array}{llll}
8.33\times 10^{\text{-2}} & -1.08\times 10^{\text{-4}} & 0 & 0 \\
-1.08\times 10^{\text{-4}} & 9.62\times 10^{\text{-6}} &
-1.19\times 10^{\text{-14}} & 0 \\
0 & -1.19\times 10^{\text{-14}} & 2.50\times 10^{\text{-21}}
& -2.01\times 10^{\text{-44}} \\
0 & 0 & -2.01\times 10^{\text{-44}} & 1.63\times 10^{\text{-67}}
\end{array}
\right),\ee
and all gauge fields are Abelian.

With the same reservations one can relate the $3\times3$  matrix
$K^{IJ}(2)$:
\be\label{5.9}K^{IJ}(2)=\left(
\begin{array}{lll}
2.99\times 10^{\text{-3}} & -1.51\times 10^{\text{-7}} & 0 \\
-1.51\times 10^{\text{-7}} & 6.58\times 10^{\text{-10}} & -5.12\times
10^{\text{-22}} \\
0 & -5.12\times 10^{\text{-22}} & 4.04\times 10^{\text{-34}}
\end{array}
\right),\ee
to grand unification theories (GUT) extended by ``high-energy gravitation'', $K^{22}(2)$, and ``high-energy cosmological constant'', $K^{33}(2)$.
The next $2\times2$  matrix $K^{IJ}(3)$:
\be\label{5.10}K^{IJ}(3)=\left(
\begin{array}{ll}
2.17\times 10^{\text{-6}} & -2.15\times 10^{\text{-12}} \\
-2.15\times 10^{\text{-12}} & 2.22\times 10^{\text{-17}}
\end{array}
\right),\ee
may be associated
with a super-unification including some version of ``high-energy gravity'', since out of five low-energy (for $e=0$) interactions there survive
only two of them which correspond to ``high-energy gravitational'', $K^{11}(3)$, and
``high-energy cosmological'', $K^{22}(3)$, coupling constants.  
In our model, it is natural associate the
$1\times1$ matrix (one rational number) 
$$K^{IJ}(4)=\left(4.48\times 10^{\text{-9}}\right)$$
with the Planck scales,  since on this level the unique interaction
(high-energy pre-image of the ``cosmological'' one) remains. It is obvious that in
order to these interrelations might have some sense, we should first
introduce some additional structures (such as metric, conformal {\it etc.}) on the seven-dimensional space-time manifold $X^4\times M^3_{+}$,
then construct over them field theories with local degrees of freedom.
However in this paper we confine ourself to the pure topological level and try to show
that the global vacuum structures already give us the hierarchy of coupling constants.

As a consequence of the above considerations
we can conjecture that in our model there exists a ``running coupling constants'' effect in its discrete mode. It may be expressed as a dependence of couplings on the discrete
energy parameter $e$. To explain this phenomenon note that the definite diagonal element of linking matrices $K^{IJ}(e)$ $(I,J = 1,\dots, 5-e)$ corresponds to each node of the splice diagram depicted in figure 7. We give these diagonal elements  $K^{II}(e)$ in Table~\ref{tab:3}, where the parameter $n$ is connected with the number $I$ of the diagonal element by
$I=n-e+1$, when $n$ runs from $e$ to 4 and $e$ is fixed.

We suppose that the identification of low-energy interactions according to the coupling constant hierarchy (described by the diagonal elements $K^{II}(0)$) may be extended along the horizontal lines ($n=$const) of Table~\ref{tab:3}. This supposition gives a possibility to say that the coupling constants $K^{II}(e)$ associated with a specific interaction depend on the discrete energy scale parameter $e$. For example, when $e$ runs from 0 to 4 the ``cosmological constant''  $K^{RR}(e)=K^{5-e,5-e}(e)$
($n=4$ and therefore $R=n-e+1=5-e$) changes according to the following sequence:
$$2.66\times 10^{-134}\rightarrow 1.63\times 10^{-67}\rightarrow 4.04\times 10^{-34}\rightarrow 2.22\times 10^{-17}\rightarrow 4.48\times 10^{-9}.$$
If we correspond the last value of the cosmological constant to the unit Planck scale, then the running cosmological constant reduced to the Planck units reads 
$$5.94\times 10^{-126}\rightarrow 3.64\times 10^{-59}\rightarrow 9.02\times 10^{-26}\rightarrow 4.96\times 10^{-9}\rightarrow 1.$$
The corresponding sequence of vacuum energy scales is
\be\label{5.11} 6.0\times 10^{-13}{\rm GeV}\rightarrow 3.0\times 10^{4}{\rm GeV}\rightarrow 6.7\times 10^{12}{\rm GeV}\rightarrow 1.0\times 10^{17}{\rm GeV}\rightarrow 1.2\times 10^{19}{\rm GeV}.\ee
(We emphasize that the Planck energy scale $1.2\times 10^{19}{\rm GeV}$ is used as a normalization and note that our model has two energy scales, associated with GUT, namely $6.7\times 10^{12}{\rm GeV}$ and $1.0\times 10^{17}{\rm GeV}$ corresponding to the matrices (\ref{5.9}) and (\ref{5.10}) respectively.)
Represented in this form, the ``running cosmological constants'' acquire the sense of the vacuum energy scales associated with the topology changes of the extra dimensional space, which induce the unification of gauge interactions.
The first term of the sequence (\ref{5.11}), corresponding to the low-energy (cosmological) vacuum density (dark energy density) , i.e. $6.0\times 10^{-13}$, is less than the empirical data, $E_{\rm vac}^{\rm empir}=(2.33\pm 1.10)\times 10^{-12}(\rm GeV)$ \cite{Bous2} only on the factor 3.9. The other terms it is possible to compare with the symmetry breaking scales, which correspond to the phase transformations of vacuum states.

\begin{table*}

\caption{\label{tab:3} Diagonal elements $K^{II}(e)$ as ``running coupling constants''.}
\begin{center}
\begin{tabular}{|c|l|l|l|l|l|l|}
\hline
Inter.&~~~~~~ $\hspace*{-0.9cm}_{n}$\hspace*{-.2cm}$\diagdown$\hspace*{-.1cm}$^{e}
$\hspace*{-.4cm} & $0$ & $1$ & $2$ & $3$ & 4 \\
\hline strong& 0 & $9.68\times 10^{-1}$ &  &  &  &   \\
\hline electr.& 1  & $7.21\times 10^{-3}$ &$8.33\times 10^{-2}$  &&& \\
\hline weak &2 & $1.76\times 10^{-12}$ &$ 9.62\times 10^{-6}$ &$2.99\times  10^{-3}$
&&\\
\hline grav.& 3 &$ 3.68\times 10^{-44}$ &$2.50\times 10^{-21}$&$6.58\times 10^{-10}$ &$2.17\times 10^{-6}$
&\\
\hline cosm.&4 &$2.66\times 10^{-134}$&$1.63\times 10^{-67}$&$4.04\times 10^{-34}$&
$2.22\times 10^{-17}$&$4.48\times 10^{-9}$\\
\hline
\end{tabular}
\end{center}
\end{table*}
Now we want to proceed to the discussion of fine tuning effect for the coupling constants as it is modeled in our scheme. 
The cosmological constant problem strongly suggest the existence of a fine-tuning mechanism, since the empirical energy density of cosmological vacuum is at least 60 orders of magnitude smaller than several theoretic contributions to it. Recall that in quantum field theory some contributions to the vacuum density are evaluated as follows (see \cite{{Rubakov}, {Bous}, {Bous2}, {Polchin}}): from the standard theory $(200 {\rm GeV})^4\approx 10^{-67}$; from the low energy supersymmetry breaking scale $(10^3 {\rm GeV})^4\approx 10^{-64}$; from grand unification schemes  $(10^{13} {\rm GeV})^4~ - ~ (10^{16} {\rm GeV})^4\approx 10^{-24}~ - ~ 10^{-12}$ (depending on a model); from quantum gravity $(10^{19} {\rm GeV})^4\approx 1$.
Within our framework an enormous fine tuning
for the cosmological constant is modeled owing to the topological properties of graph manifolds under consideration.

Note that in our model the  ``running cosmological constant'' (or the sequence of vacuum energy scales) is associated with the last   
diagonal elements $K^{RR}(e)=K^{5-e,5-e}(e)$ of rational linking matrices of the graph manifolds $\{M^3_{+}(e)|e\in \overline{0,4}\}$ and thus undergoes a change when the topology of extra-dimensional space is transformed. Therefore the cosmological constant, understood as vacuum energy density,  depends on the discrete energy parameter $e$, as is shown in the last line of the Table \ref{tab:3}. We give now the elements of this line in the form that demonstrates the fine tuning of the cosmological constant corresponding to the different levels marked by the energy parameter $e$. 

Low energy cosmological constant ($e=0$) reads (see formula (\ref{5.6}))
\be\label{5.12}K^{55}(0)=\sum_{i=1}^3 \frac{b_i(0)}{a_i(0)},\ee
where
$$\begin{array}{l}\frac{b_1(0)}{a_1(0)}=\frac{-5}{19};\\ \frac{b_2(0)}{a_2(0)}=\frac{84986185048470683619597440588949913929045173055232557772918123429}{322947503184188597754470274238009672930371657609883719537088869030};\\
\frac{b_3(0)}{a_3(0)}=\\
\frac{-5224868486304546518657}{
32059806410238279718139966062484464209573437243152889136180581555726013473961031111880160}.\end{array}$$
The fine tuning manifests itself in following manner.
The first two terms have order $10^{-1}$:
$$\begin{array}{l}-0.26315789473684210526315789473684210526315789473684210526315789473\,6...\\
~~0.26315789473684210526315789473684210526315789473684210526315789473\,7... \end{array}$$
and their sum is 
$$~1.629725525927064843047085356241132473717675047642589547871480144872\,7...\times 10^{-67}.$$
The third term is
$$-1.629725525927064843047085356241132473717675047642589547871480144872\,4...\times 10^{-67}.$$
The final result is
\be\label{5.13}K^{55}(0)=\sum_{i=1}^3 \frac{b_i(0)}{a_i(0)}=2. 656005289...\times 10^{-134}.\ee
The fine tuning for the cosmological constant is really wonderful: two terms of order $10^{-1}$ and one of order $10^{-67}$ have canceled mutually with the accuracy ~$10^{-134}$.

The ``cosmological'' constant (vacuum energy density) corresponding to the first ($e=1$) level of ``unification'' (speculatively interpreted before the formula (\ref{5.8}) as the level of extended standard model) reads (see formula (\ref{5.4}))
\be\label{5.14}K^{44}(1)=\sum_{i=1}^3 \frac{b_i(1)}{a_i(1)},\ee
where
$$\begin{array}{l}\frac{b_1(1)}{a_1(1)}=\frac{-5}{19};\\
\frac{b_2(1)}{a_2(1)}=\frac{34308803181407992475839813298929}{130373452089350371408191290535930};\\
\frac{b_3(1)}{a_3(1)}= \frac{-20047766957}{49660235112471118762743483357286701277802160}. \end{array}$$
The fine tuning manifests itself in the same manner as in the previous case.
The first two terms have order $10^{-1}$:
$$\begin{array}{l}\frac{b_1(1)}{a_1(1)}=-0.263157894736842105263157894736842\,1...\\
\frac{b_2(0)}{a_2(0)}=~0.263157894736842105263157894736842\,5...\end{array}$$
and their sum is 
$$~4.03698591269162697216671705018608\,3...\times 10^{-34}.$$
The third term is
$$-4.03698591269162697216671705018608\,1...\times 10^{-34}.$$
The final result demonstrate once again the great fine adjustment:
\be\label{5.15}K^{44}(1)=\sum_{i=1}^3 \frac{b_i(1)}{a_i(1)}=1.62972552\,6983...\times 10^{-67}.\ee 

We may observe, that the value of the coupling constant
$K^{44}(1)$ and the absolute value of the contribution \be\label{5.16}\frac{b_3(0)}{a_3(0)}=-1.62972552\,5927...\times 10^{-67}\ee in the low-energy cosmological constant $K^{55}(0)$ match up to the eighth order. It may appear an attractive illusion  that this contribution in the low energy cosmological constant originates from the vacuum energy density of the extended standard model and is fine tuned by other two contributions $\frac{b_1(0)}{a_1(0)},~~\frac{b_2(0)}{a_2(0)}$ . But  this hope almost fades out, when we consider these numbers in the fraction representation:
$$\begin{array}{l}K^{44}(1)=\frac{26869}{164868252691154276926871239187866021757572803
005236098566405242449679440},\\
\frac{b_3(0)}{a_3(0)}=\\
\frac{-5224868486304546518657}{32059806410 238279718139966062484464209573437243152889136180581555726013473961031111880160}. \end{array}$$
In either case this coincidence is more than impressive.

The ``cosmological'' constant corresponding to the second ($e=2$) level of ``unification'' (speculatively interpreted before the formula (\ref{5.9})
as the level of GUT model extended by ``high-energy gravitation'' and ``high-energy cosmology'' interactions) reads (see formula (\ref{5.3}))
\be\label{5.17}K^{33}(2)=\sum_{i=1}^3 \frac{b_i(2)}{a_i(2)},\ee
where
$$
\begin{array}{l}\frac{b_1(2)}{a_1(2)}=\frac{-5}{19}=-0.\overline{263157894736842105};\\
\frac{b_2(2)}{a_2(2)}=\frac{689341120054429}{2619496 256206830}=
0.2631578947368421\,2...\\
\frac{b_3(2)}{a_3(2)}=\frac{-39257}{1953837726068319020160}=
-2.0092252020845316\,1...\times 10^{-17}.\end{array}$$
The sum of the first two terms is
$$\begin{array}{l}\frac{b_1(2)}{a_1(2)}+\frac{b_2(2)}{a_2(2)}=\frac{1}{49770428867929770}=
2.0092252020845316\,5\times 10^{-17}.\end{array}$$
Finally we obtain the fine tuning on the level $e=2$:
\be\label{5.18}K^{33}(2)=\sum_{i=1}^3 \frac{b_i(2)}{a_i(2)}=4.0383227651... \times10^{-34}.\ee 
Note that in this case once again, we have the coincidence of vacuum energy density $K^{33}(2)$ with the absolute value of third  contribution in $K^{44}(1)$, namely,
$\frac{b_3(1)}{a_3(1)}=-4.0369859126...\times10^{-34},$ but only up to second order.

All information about the fine tuning for the ``cosmological'' constant $K^{22}(3)$, corresponding to the third ($e=3$) level of ``unification'' (an intermediate level, between the extended GUT unification and Planck scale of energy), already was given in the formula (\ref{5.2}).The ``cosmological'' constant or vacuum energy density corresponding to the fourth ($e=4$), the last in this framework, level of ``unification'' (the Planck scale of energy) yet is represented in (\ref{5.1}) and coincide with our preliminary example (\ref{4.4}). 

The following observations are important:

1. In this section we have used the rather simple notations $K^{RR}(e)=\sum_{i=1}^3 \frac{b_i(e)}{a_i(e)}$. In order to bring these notations into conformity with the preceding ones, which we used in formula (\ref{2.45}) for the last diagonal element of linking matrix, it is necessary to add the energy parameter $e$ in all terms of (\ref{2.45}). Thus we obtain 
$$K^{RR}(e)=-\left(\frac{{q*}^{R-1}(e)}{p^{R-1}(e)}+\frac{b^R(e)}{a^R(e)}+\frac{b^{R+1}(e)}{a^{R+1}(e)}\right).$$ Then the accordance is the following:
$$\frac{b_1(e)}{a_1(e)}=-\frac{b^R(e)}{a^R(e)};~  \frac{b_2(e)}{a_2(e)}=-\frac{b^{R+1}(e)}{a^{R+1}(e)};~  \frac{b_3(e)}{a_3(e)}=-\frac{{q*}^{R-1}(e)}{p^{R-1}(e)}.$$

2. Each ``cosmological'' constant has the same contribution $-\frac{5}{19}$ with the absolute value of order $10^{-1}$ and an other term with the same order, but positive, which compensates the the first one with great precision. The remainder  has the same order as the third term and they are compensated each other with enormous exactness.  The formula (\ref{5.1}) presents the unique exception out of this mechanism and describe more weak fine tuning.

3. The coupling constant for $(e+1)$-level have the same order as the third term in the coupling constant expression for $e$-level for all $e\in \overline{0,3}$.

4. In our model the fine tuning effect takes place for coupling constants of all other gauge interactions (see diagonal elements in the formulas (\ref{5.2})-(\ref{5.4}) and (\ref{5.6})). This fact may be useful for the resolution of other fine-tuning problems such as a reconsideration of the fine-tuning problem in low-energy SUSY models, motivated by the recent observation of the relatively heavy Higgs boson \cite{1301.1137}.

5. All the expressions for coupling constants have the same form (see the formulas (\ref{2.41}), (\ref{2.44}) and (\ref{2.45})) of type 
$$K^{II}= -\left(\frac{{q*}^{I-1}}{p^{I-1}}+\frac{q^I}{p^I}+\frac{b^I}{a^I}\right),$$
and contain always three terms which are fine tuned. This is a model dependent effect and is connected with the fact that we use exclusively the Breiskorn homology spheres (which have only three exceptional orbits) for constructing the graph manifolds shown in Fig. 3. The Breiskorn homology spheres form the simplest class of Seifert fibered homology spheres, which in general case have an arbitrary finite number of exceptional orbits (see Appendix). Using the general Seifert fibered homology spheres in graph manifold construction we can obtain  formulas of the same type for the coupling constants, but with arbitrary number of terms.

6. It is possible that the splitting off the set of lens spaces (baby universes in internal space) $L(p^K_s,q^K_s)$ and $L(a^L_t,b^L_t)$ is the physical reason due to which the fine tuning for all coupling constants takes place, as was described in the neighborhood of the formula (\ref{2.56}). Here $\frac{p^K_s}{q^K_s}=-[e_s^K,\dots, e^K_{n_K}]$,~~ $s=1,\dots,n_K$ and $\frac{a^L_t}{b^L_t}=-[\epsilon_t^L,\dots, \epsilon_{m_L}^L]$,~~ $t=1,\dots,m_L$, see  the end of subsection 2.2.

\section{Conclusions}\label{s6}
\setcounter{equation}{0}
We consider as the main result of our paper the following one: by means of using the ge\-neral properties of topological invariants (rational linking matrices) of graph manifolds we succeeded in modeling the hierarchy of low energy of coupling constants as well as their fine tuning. Our approach to these problems is not dynamical since the low energy coupling constants are extracted formally from the three additional dimensions according to the Kaluza-Klein procedure. We have not proposed dynamical mechanisms to choose the internal space. This situation is reminiscent of that which occurs in the gauge theory of topological fluids \cite{WenObzor}. There exists the procedure for the formation of topological fluids ground states by means of sequential condensations of topological defects in the Julia-Toulouse approach, but the conditions and the dynamical mechanism for the condensation are not discussed \cite{{JulToul}, {QuevTrug}, {DiamantI}}. As in the case of BF theory in terms of Julia-Toulouse condensation \cite{{BrazJHEP}, {Braz1202.3798}}, we have constructed  an effective field theory with vacuum coupling constants (induced by the topological structure of internal spaces) for low energy excitations described by the collection of gauge fields. The set of internal three dimensional graph manifolds has been built in Section 4 according to the rather complex and unusual (for physicists) algorithm, but all used topological operations are well defined within the framework of topology of three-manifolds. Because of the unusual simulation method we used in an attempt to solve the hierarchy and fine tuning problems, we would like to summarize the results of our paper also on an unusual way. 

It is appropriate to start from a quotation attributed to Leopold Kronecker, who once wrote: ``God made the {\it integers}; all else is the work of man''. The first essential step in our model is the introduction of the {\it integer} linking matrix $Q^{AB}(\Gamma_p)$ of graph manifold $M(\Gamma_p)$ (see, formula (\ref{2.46})).  
This matrix reflects the structure of the plumbing graph $\Gamma_p$ and consists of
the tridiagonal submatrices of type 
$$K^{ab}\left(\underline{e},I\right) =\left( 
\begin{array}{cccc}
e_{1}^{I} & -1 &  &  \\ 
-1 & \ddots  &  &  \\ 
&  & \ddots  & -1 \\ 
&  & -1 & e_{n_{I}}^{I}
\end{array}
\right) 
~~\text{ and }~~
K^{\alpha \beta }\left(\underline{\epsilon},I\right) =\left( 
\begin{array}{cccc}
\epsilon _{m_{I}}^{I} & -1 &  &  \\ 
-1 & \ddots  &  &  \\ 
&  & \ddots  & -1 \\ 
&  & -1 & \epsilon _{1}^{I}
\end{array}
\right)$$
which are associated with the low-energy physics of hierarchical topological fluids with $n_I$ and $m_I$ levels respectively \cite{{WenZee}, {FujitaLi}}. These topological fluids are united according to the structure of the graph $\Gamma_p$  and their main characteristics  ({\it i.e.} filling factors determined by the formulas (\ref{2.47}) - (\ref{2.50})) form diagonal elements of the {\it rational} linking matrices $K^{IJ}$ of internal spaces in seven dimensional Kaluza-Klein approach (see equations (\ref{2.51}) and (\ref{2.52})). These rational linking matrices are interpreted, according to their position in BF-action (\ref{3.16}), as coupling constants matrices (up to the dimensionless scale factor $k$). The examples constructed in the sections 4 and 5 show that the rational linking matrix $K^{IJ}(0)$ contains (in its diagonal elements) the hierarchy of coupling constants of five fundamental low-energy interactions of the nature (see boldface numbers in the matrix (\ref{5.7})). But, on the other hand, these matrices can be obtained by the Gauss process of partial diagonalization of the {\it integer} linking matrix $Q^{AB}(\Gamma_p)$, as it was done by W. Neumann in \cite{Neum77}. He called them reduced plumbing matrices $K^{IJ}_{\rm reduced}$. We have demonstrated in subsection 2.2 that these matrices coincide with the rational linking matrices; and (if applying the notions of sections 4 and 5) depend on the discrete energy parameter $e$,  {\it i.e.} $K^{IJ}_{\rm reduced}(e)=K^{IJ}(e)$. In addition to the matrix  $K^{IJ}_{\rm reduced}(e)$ the partially diagonalized matrix 
\be \label{6.1} Q^{AB}_{\rm{part.diag}}(\Gamma_p(e)) = K^{IJ}_{\rm{reduced}}(e)\oplus D^{MN}(e)\ee (see equation (\ref{2.54})) contains the diagonal part  $D^{MN}(e)$. (Here the graph $\Gamma_p (e)$ and all matrices depend on the parameter $e$ that is clear from the Fig. 7.) According to our crude estimates rank$D^{MN}(0)\approx 4.0\times 10^{20}$. Each element of $D^{MN}(e)$ is represented by continued fraction of type 
\be \label{6.2}-\frac{p^K_s(e)}{q^K_s(e)}=[e_s^K,\dots, e^K_{n_K}](e),~~ {\rm and}~~  
-\frac{a^L_t(e)}{b^L_t(e)}=[\epsilon_t^L,\dots, \epsilon_{m_L}^L](e),\ee
see formulas (\ref{2.55}). All these fractions depend on discrete energy parameter $e$. Each of the fractions is the rational linking $(1\times 1)$-matrix of certain lens space, such as $L(p^K_s(e), q^K_s(e))$ or $L(a^L_t(e), b^L_t(e))$. We consider the set of these lens spaces as a collection of internal spaces of disjoint (according to the Fig. 6) seven-dimensional 
``baby-universes'' of type $X^4\times L(p^K_s(e), q^K_s(e))$ or $X^4\times L(a^L_t(e), b^L_t(e))$. After the dimensional reduction down to four-dimensional space-time each $X^4$  gets one gauge interaction with the coupling constant $-\frac{p^K_s(e)}{q^K_s(e)}$ or $-\frac{a^L_t(e)}{b^L_t(e)}$ respectively. Therefore
we must endow these disjoint space-times $X^4$ with the same indexes that the internal lens space has, that is,  $X^{4,K}_s(e)$ or $X^{4,L}_t(e)$. 
Thus, now we can rewrite the formula (\ref{2.56}) for the total seven-dimensional manifold which corresponds to the plumbing graphs shown in Figs.~\ref{f5} and~\ref{f6} and to the partially diagonalized matrix (\ref{6.1}) as follows  
\be\label{6.3}X^4(e)\times M^3_{+}(e) \bigsqcup_{K=1}^R \bigsqcup_{s=1}^{n_K}X^{4,K}_s(e)\times L(p^K_s(e),q^K_s(e))\bigsqcup_{L=0}^{R+1}\bigsqcup_{t=1}^{m_L}X^{4,L}_t(e)\times L(a^L_t(e),b^L_t(e)).\ee
We can interpret this situation as follows: In order to obtain a large universe with hierarchical set of {\it rational} coupling constants $K^{II}(e)$, an enormous (but finite) number of
``baby-universes'' $\{X^{4,K}s(e), X^{4,L}_t(e)\}$ have to be ``chipped off''. Each of these ``baby-universes'' has only one gauge interaction with {\it rational} coupling constant $-\frac{p^K_s(e)}{q^K_s(e)}$ or $-\frac{a^L_t(e)}{b^L_t(e)}$ with absolute value greater than 1, thus the scales of all of them would be of Planck or sub-Planck orders. 

God made an {\it integer} linking matrix $Q$; all else is our {\it rational} but dubious work.

Actually, we introduce in the algorithm for calculation of coupling constants matrices only first nine prime numbers $p_i$: $2,3,5,\dots 23$ and 1. We use these numbers and their products, $q_{2n-1}=p_1\cdots p_{2n-1}$ to define the subsequence    
$$\left\{\Sigma_{+}(p_{2n},p_{2n+1},q_{2n-1})|n\in\overline{1,4}\right\}$$
of the primary sequence (\ref{4.3}) of Breiskorn homology spheres. Further, the algorithm described in sections 4 and 5 gives us, as five low-energy coupling constants, the diagonal elements of the matrix (\ref{5.7}). These five numbers are rational ones and are presented in formulas (\ref{5.6}).  It is possible to choose instead of the primary set of numbers, $1,2,3,5,\dots 23$, any (arbitrary) ordered collection of  ten positive pairwise relatively prime integers $n_1\dots n_{10}$ (or 1 and nine relatively prime numbers). This choice leads to the change of the primary subsequence of Breiskorn homology spheres, and our algorithm will give other set of coupling constants with rather different hierarchy. Thus we obtain an infinite ``landscape'' of universes with very different hierarchy of coupling constants, but that's wonderful that the hierarchy typical for our Universe is determined by the natural sequence of {\it first nine prime} numbers (without any anthropic principle).

May be: God made {\it prime} numbers; all else is the work of man.

\section*{Acknowledgments}
We would like to thank Chris Beasley for his kind clarification of the issue about line bundles and characteristic classes on orbifolds.

\appendix

\section{Seifert Fibered Homology Spheres with Unnormalized Seifert Invariants }
\label{sec:A}
\setcounter{equation}{0}

In this paper we use Brieskorn homology spheres (Bh-spheres) as the constructive blocks of graph manifolds. Bh-spheres are a special case of Seifert fibered homology spheres. It is appropriate to reproduce here the basic definitions of these classical objects, following to \cite{{EisNeum}, {SavB2}}.

Let $a_1,\dots, a_n$ be pairwise coprime integers with $n\geq 3$ and each $a_k\geq 2$. Let $C=(c^{ik})$ be an $(n-2)\times n$-matrix of complex numbers such that each of the maximal minors of $C$ is non-zero. A {\it Seifert fibered homology sphere} (Sfh-sphere) is defined as the link of singularity at the origin
\be\label{A.1} \Sigma(a_1,\dots, a_n)=\left\{\sum_{k=1}^n c^{ik}z_k^{a_k},~i\in\overline{1,(n-2)}=0\right\}\cap S^{2n-1}.\ee
It is a smooth  three dimensional manifold with induced orientation which does not depend on the matrix $C$ up to homeomorphism. The homology groups with integer coefficients of a Sfh-sphere are isomorphic to ones of $S^3$, thus each Sfh-sphere is an integer homology sphere ($\mathbb Z$-homology sphere). The case of minimal $n$, {\it i.e.} $n=3$, gives us Bh-spheres (see formula (\ref{4.1})).

Sfh-sphere $\Sigma(a_1,\dots, a_n)$ can be described topologically as follows. Let $F_0=S^2\setminus {\rm int}(D^2_1\cup\dots\cup D^2_n)$ be the $n$-fold punctured two-sphere and consider an $S^1$-bundle $\Sigma_0\rightarrow F_0$ with an integer Euler number $b$ and with fixed trivialization over $\partial F_0$. The boundary of $\Sigma_0$ consists of $n$ tori $(\partial D^2_k)\times S^1$. Given $n$ pairs of relatively prime integers, $(a_k,b_k)$, paste $n$ solid tori $D^2_k\times S^1$ into $\Sigma_0$ in such a way that the homology class $a_k(S^1\times \{1\})+b_k(\{1\}\times S^1)$ in the $k$-th boundary component of $\Sigma_0$ is null-homologous in $D^2_k\times S^1$ after pasting. The integers $b,~(a_1,b_1),\dots, (a_n,b_n)$ are called Seifert invariants of $\Sigma(a_1,\dots, a_n)$. The obtained manifold is $\mathbb Z$-homology sphere if and only if and only if $\sum (b_k/a_k)-b=\pm 1/a$, where $a=a_1\cdots a_n$. We fix the orientation by choosing $-1$, so that we obtain the rational Euler number of Sfh-sphere, also known as the Chern class of the line V-bundle associated with the $S^1$-bundle over two-dimensional orbifold with $n$ conic singularities \cite{{FurutaSteer},{GriTan}} 
\be\label{A.2}e(\Sigma)=\sum_{k=1}^n\frac{b_k}{a_k}-b=-\frac{1}{a}.\ee
Rewriting this formula in manner $a~\sum (b_k/a_k)=a~b-1$ and reducing it mo\-dulo $a_k$, we see that the $b_k$ are uniquely determined modulo $a_k$ for each $k$. It was shown that two $\mathbb Z$-homology spheres with Seifert invariants $b,~(a_1,b_1),\dots, (a_n,b_n)$ and $b',~(a_1,b'_1),\dots, (a_n,b'_n)$ are orientation preserving homeomorphic if and only if $b'_k=b_k$ modulo $a_k$ for all $k$. Thus the invariants $a_k$ together with the equation (\ref{A.2}) determine a unique Sfh-sphere $\Sigma(a_1,\dots, a_n)$. Also note that one can always choose the invariants $b_k$ so that $b=0$. In this case the Seifert invariants are called {\it unnormalized}, and the $S^1$-bundle $\Sigma_0\rightarrow F_0$ is trivial. We use the unnormalized Seifert invariants everywhere in our paper.  

All $\mathbb{Z}$-homology spheres can be constructed from Bh-spheres according to plumbing  diagrams by the plumbing (splicing) operation.
This operation is defined for any Sfh-sphere
as follows: First we define \cite{EisNeum} a Seifert link
(not to confuse with a link of singularity) as a
pair $(\Sigma,S)=(\Sigma,S_1\bigcup\cdots \bigcup S_m)$ consisting
of  an oriented Sfh-homology sphere $\Sigma$ and a collection
$S$ of Seifert fibers (exceptional or regular) $S_1,\dots,S_m$ in
$\Sigma$. Note that the links $(S^3,S)$ where $S^3$ is an ordinary
three-sphere, are also allowed. Let $(\Sigma,S)$ and
$(\Sigma',S')$ be links and choose components $S_i\in S$ and
$S'_j\in S'$. Let also $N(S_i)$ and $N(S'_j)$ be their tubular
neighbourhoods, while $m,l\subset\partial N(S_i)$ and
$m',l'\subset\partial N(S'_j)$ be standard meridians and
longitudes. The manifold
$\Sigma''=(\Sigma\setminus\textnormal{int}N(S_i))
\bigcup(\Sigma'\setminus\textnormal{int}N(S'_j))$ obtained by
pasting along the torus boundaries by matching $m$ to $l'$ and
$m'$ to $l$, is a ${\mathbb Z}$-homology sphere. 
We shall use the standard notation
$\Sigma''= \Sigma\frac{~}{S_i ~ S'_j}\Sigma'$ or simply $\Sigma''=
\Sigma\frac{~ ~}{~ ~ ~}\Sigma'$.
All ${\mathbb{Z}}$-homology spheres have been classified in
\cite{EisNeum} by their plumbing diagrams.

\end{document}